\documentstyle[amsmath,epsf,psfig,epsfig]{mn}

\newcommand{\beqa}{\begin{eqnarray}}
\newcommand{\eeqa}{\end{eqnarray}}
\newcommand{\f}{\frac}

\def\gsim {\lower .1ex\hbox{\rlap{\raise .6ex\hbox{\hskip .3ex
        {\ifmmode{\scriptscriptstyle >}\else
                {$\scriptscriptstyle >$}\fi}}}
        \kern -.4ex{\ifmmode{\scriptscriptstyle \sim}\else
                {$\scriptscriptstyle\sim$}\fi}}}
\def\lsim {\lower .1ex\hbox{\rlap{\raise .6ex\hbox{\hskip .3ex
        {\ifmmode{\scriptscriptstyle <}\else
                {$\scriptscriptstyle <$}\fi}}}
        \kern -.4ex{\ifmmode{\scriptscriptstyle \sim}\else
                {$\scriptscriptstyle\sim$}\fi}}}
\def\kmsmpc{\,{\rm km\,s^{-1}\,Mpc^{-1}}}
\def\msun{\,{\rm M_\odot}}
\def\lta{\mathrel{\spose{\lower 3pt\hbox{$\mathchar"218$}}
     \raise 2.0pt\hbox{$\mathchar"13C$}}}
\def\gta{\mathrel{\spose{\lower 3pt\hbox{$\mathchar"218$}}
          \raise 2.0pt\hbox{$\mathchar"13E$}}}

\def\etal{et al. }
\def\eg{e.g.}

\def\pa{\partial}
\def\prop{\propto}

\def\ltsima{$\; \buildrel < \over \sim \;$}
\def\lsim{\lower.5ex\hbox{\ltsima}}
\def\gtsima{$\; \buildrel > \over \sim \;$}
\def\gsim{\lower.5ex\hbox{\gtsima}}

\def\la{\langle}
\def\ra{\rangle}

\def\se#1{\S\ref{sec:#1}}
\def\fig#1{Fig.~\ref{fig:#1}}
\def\equ#1{Eq.~(\ref{eq:#1})}
\def\ifm#1{\relax\ifmmode#1\else$\mathsurround=0pt #1$\fi}

\def\hmpc{\,h\ifm{^{-1}}{\rm Mpc}}
\def\hkpc{\,h\ifm{^{-1}}{\rm kpc}}
\def\solar{\ifmmode_{\mathord\odot}\;\else$_{\mathord\odot}\;$\fi}
\def\msun{{\rm M}_{\solar}}
\def\omm{\Omega_{\rm m}}

\def\am{{\bf L}}
\def\bfx{{\bf x}}
\def\bfr{{\bf r}}
\def\bfv{{\bf v}}
\def\bfq{{\bf q}}
\def\bfS{{\bf S}}

\def\bfi{{\bf i}}
\def\bft{{\bf t}}
\def\bfL{{\bf L}}
\def\D{{\cal D}}

\begin{document}

\title[Shear vs. Inertia and Proto-haloes]
      {Testing Tidal-Torque Theory: II. Alignment of Inertia and Shear
       and the Characteristics of Proto-haloes}

\author[C. Porciani, A. Dekel \& Y. Hoffman]
{Cristiano Porciani$^{1,2}$, Avishai Dekel$^1$ and Yehuda Hoffman$^1$\\
$^1$Racah Institute of Physics, The  Hebrew University,
Jerusalem 91904, Israel\\
$^2$Institute of Astronomy, University of Cambridge,
Madingley Road, Cambridge CB3-OHA, UK}

\maketitle
\begin{abstract}

We investigate the cross-talk between the two key components of tidal-torque
theory, the inertia ($I$) and shear ($T$) tensors, using a cosmological
$N$-body simulation with thousands of well-resolved haloes.  We find that the
principal axes of $I$ and $T$ are {\it strongly aligned}, even though $I$
characterizes the proto-halo locally while $T$ is determined by the large-scale
structure.  Thus, the resultant galactic spin, which plays a key role in galaxy
formation, is only a residual due to $\sim$10 per cent deviations from perfect
alignment of $T$ and $I$.  The $T$-$I$ correlation induces a weak tendency for
the proto-halo spin to be perpendicular to the major 
axes of $T$ and $I$, 
but this
correlation is erased by non-linear effects at late times, making the observed
spins poor indicators of the initial shear field.

However, the $T$-$I$ correlation implies that the shear tensor can be used for
identifying the 
{\it positions} and 
{\it boundaries} of {\it proto-haloes} in cosmological initial conditions
--- a missing piece in galaxy formation theory.
The typical configuration is of a prolate proto-halo lying perpendicular to a
large-scale high-density ridge, with the surrounding voids inducing compression
along the major and intermediate inertia axes of the proto-halo.
This leads to a transient sub-halo filament along the large-scale ridge, 
whose sub-clumps then flow along the filament and merge into the final halo. 

The centres of proto-haloes tend to lie in $\sim 1\sigma$ over-density regions,
but their association with linear density maxima smoothed on galactic scales is
vague: only $\sim 40$ per cent of the proto-haloes contain peaks within
them.  Several other characteristics distinguish proto-haloes from density
peaks, e.g., they tend to compress along two principal axes while many peaks
compress along three axes.

\end{abstract}


\section{Introduction}
\label{sec:intro}

The angular-momentum properties of a galaxy system
must have had a crucial role in determining its evolution and final type.
A long history of research, involving Hoyle (1953) and Peebles (1969),
led to a `standard' theory for the origin of angular momentum in the
cosmological framework of hierarchical structure formation, the tidal-torque
theory (hereafter TTT), as put together by Doroshkevich (1970) and
White (1984).
Special interest in the subject has been revived recently because of an
`angular-momentum crisis', arising from cosmological simulations of
galaxy formation which seem to yield
luminous galaxies that are significantly smaller, and of much less
angular momentum than observed disc galaxies
(Navarro, Frenk \& White 1995; Navarro \& Steinmetz 1997, 2000).
Another current motivation comes from weak lensing studies,
where the intrinsic distribution and alignment of
galaxy shapes, which may be derived from TTT, play an important
role in interpreting the signal
(Croft \& Metzler 2000; Heavens, Refregier \& Heymans 2000;
Catelan, Kamionkowski \& Blandford 2001; Catelan \& Porciani 2001; 
Crittenden \etal 2001).

These add a timely aspect to the motivation for revisiting
the classical problem of angular momentum, in an attempt to sharpen
and deepen our understanding of its various components and how they work
together. In particular, this effort should start from the acquisition of
angular momentum by dark-matter haloes, despite the misleading
apparent impression that this part of the theory is fairly well understood.
The basic notion of TTT is that most of the angular momentum is being
gained gradually by proto-haloes in the linear regime of density fluctuations
growth, due to tidal torques from neighboring fluctuations.
This process is expected to continue as long as the proto-halo is expanding.
Once it decouples from the expanding background and turns around to non-linear
collapse and virialization, only little angular momentum is expected to be
tidally exchanged between haloes.
It is commonly assumed that the baryonic material, which in general
follows the dark-matter distribution inside each proto-halo, gains a similar
specific angular momentum and carries it along when it contracts
to form a luminous galaxy at the halo centre.
This should allow us to predict galactic spins using the approximate
but powerful analytic tools of quasi-linear theory of gravitational
instability.
Even if angular momentum is transferred from the gas to the dark matter
during disc formation, the initial set-up of angular-momentum distribution
in the halo is an important ingredient in the process.

In a series of papers, we evaluate the performance of the TTT approximation,
and trace the roles of its various ingredients and the cross-talk between them.
We do it using a cosmological $N$-body simulation with $\sim 7300$
well-resolved haloes.
We find that some of the basic ingredients of standard TTT
involve certain unjustified assumptions and that our understanding of
the theory is not full.  These papers represent attempts to
clarify some of these controversial issues.
In Paper I (Porciani, Dekel \& Hoffman 2001),
we evaluate how well does the approximation predict the final
spin of a halo, given full knowledge of the corresponding
initial proto-halo and the cosmological realization.
In the present paper (Paper II), we attempt
a deeper level of understanding of the origin of halo angular momentum,
by investigating the relation between the different components of TTT.
We find that this study connects to the fundamental open question of how to
identify a proto-halo in a given realization of initial conditions,
and how proto-haloes evolve via collapse and mergers to the present virialized
haloes. While the current discussion of proto-haloes and their evolution is 
limited to first results, a more detailed analysis will be reported in future 
papers.
In Paper III (Porciani \& Dekel in preparation), the standard
scaling relation of TTT is revised, based on our finding that the shear
tensor is weakly correlated 
with the density at the proto-halo centre of mass.
This scaling relation is used to
predict the typical angular-momentum profile of haloes (Dekel \etal 2001;
Bullock \etal 2001b), and to provide a simple way to incorporate angular
momentum in semi-analytic models of galaxy formation (Maller, Dekel \&
Somerville 2002).

According to TTT, the angular momentum of a halo is due to the
cross-talk between two key players: the inertia tensor $I$, describing
the quadrupole structure of the proto-halo, and the shear tensor $T$,
representing the external tidal field exerting the torque.
There is a controversy in the literature regarding the correlation between
these two components, which makes a big difference in the outcome.
In applications of TTT, it has commonly been assumed that $I$ and $T$ are
uncorrelated (e.g.  Hoffman 1986a,b; Heavens \& Peacock 1988;
Steinmetz \& Bartelmann 1995; Catelan \& Theuns 1996),
based on the argument that the former
is local and the latter must be dominated by external sources.
Lee \& Pen (2000, hereafter LP00),
based on simulations with limited resolution,
have raised some doubts concerning this assumption.
Other applications have made the assumption that the present-day halo
angular momentum correlates with the shear tensor of the initial conditions
(LP00; Pen \etal 2000; Crittenden \etal 2001).
This involves assuming both a strong correlation at the initial conditions
and that this correlation survives the non-linear evolution at late times.
In this paper, we investigate the validity of these assumptions, and find
surprising results that shed new light on the basic understanding of TTT.

One of our pleasant surprises is that 
a high degree of correlation between
$T$ and $I$ leads to progress in a more general problem, that of identifying
proto-haloes in the initial conditions.  Despite the impressive progress
made in analysing the statistics of peaks in Gaussian random density
fields (Bardeen \etal 1986; Hoffman 1986b;
Bond \& Myers 1996), given a realization of
such initial conditions we do not have a successful recipe for identifying
the proto-halo centers and 
the Lagrangian region about 
them 
which will end up in the final
virialized halo.  
We find that proto-haloes are only vaguely associated with density peaks. 
Straightforward ideas
for identifying the boundaries, 
such as involving iso-density
or iso-potential contours
(Bardeen \etal 1986; Hoffman 1986b, 1988b; Heavens \& Peacock 1988;
Catelan \& Theuns 1996), do not provide a successful algorithm
(van de Weygaert \& Babul 1994; Bond \& Myers 1996).
These are key missing ingredients 
in a full theory for the origin of angular
momentum as well as for more general aspects of galaxy formation theory.
We make here the first steps towards an algorithm that may provide
the missing 
pieces. 

The outline of this paper is as follows:
In \se{ttt} we summarize the basics of linear tidal-torque theory.
In \se{simu} we describe the simulation and halo finder,
and the implementation of TTT to proto-haloes.
In \se{T-I} we address the correlation between the shear field and the inertia
  tensor of the proto-halo.
In \se{gamma} we show examples of the $T-I$ correlation, and address
  the implications on the characteristics of proto-halo regions
  and how they evolve. 
In \se{properties} we refer to other properties of proto-haloes.
In \se{T-L} we address the correlation between the shear field
  and the spin direction.
In \se{conc} we discuss our results and conclude.

\section{Tidal-Torque Theory}
\label{sec:ttt}

Here is a brief summary of the relevant basics of tidal-torque theory,
which is described in some more detail in Paper I.

The framework is the standard FRW cosmology in the matter-dominated era,
with small density fluctuations that grow by gravitational instability.
Given a proto-halo, a patch of matter occupying an Eulerian volume $\gamma$
that is destined to end up in a virialized halo, the goal is to compute
its angular momentum about the centre of mass, to the lowest non-vanishing
order in perturbation theory.  The angular momentum at time $t$ is
\begin{equation}
\am (t)=\int_{\gamma} \rho(\bfr,t) \left[ \bfr (t)-\bfr_{\rm cm}(t) \right]
\times \left[ \bfv (t)-\bfv_{\rm cm}(t) \right] d^3r \, ,
\label{eq:Lprop}
\end{equation}
where $\bfr$ and $\bfv$ are the position and peculiar velocity,
and the subscript cm denotes centre-of-mass quantities.
Then, in comoving units $\bfx=\bfr/a(t)$,
\begin{equation}
\am(t)=\bar{\rho}(t) a^5(t) \int_{\gamma}
\left[ 1+ \delta(\bfx,t) \right]
\left[ \bfx (t)-\bfx_{\rm cm}(t) \right]
\times \dot{\bfx}(t) \, d^3x \, ,
\label{eq:Lcomove}
\end{equation}
where $\delta(\bfx,t)$ is the density fluctuation field
relative to the average density $\bar \rho(t)$,
$a(t)$ is the universal expansion factor,
and the term proportional to $\bfv_{\rm cm}$ vanished.
A dot denotes a derivative with respect to cosmic time $t$.

The comoving Eulerian position of each fluid element
is given by its initial, Lagrangian position $\bfq$ plus a displacement:
$\bfx=\bfq+\bfS(\bfq,t)$.
When fluctuations are sufficiently small, or when the flow is properly
smoothed, the mapping $\bfq \to \bfx$ is reversible such that the flow
is {\it laminar}.  Then the Jacobian determinant $J=|| \pa \bfx/\pa \bfq ||$
does not vanish, and the continuity equation implies
$1+\delta[\bfx(\bfq,t)]=J^{-1}(\bfq,t)$.
Substituting in \equ{Lcomove} one obtains
\begin{equation}
\am (t)=a^2(t)\,\bar{\rho}_0 a^3_0 \int_{\Gamma}
\left[\bfq-\bar{\bfq}+\bfS(\bfq,t)-\bar{\bfS}\right]
\times \dot{\bfS}(\bfq,t) \,d^3q\, ,
\label{eq:Llag}
\end{equation}
where barred quantities are averages over $\bfq$ in $\Gamma$,
the Lagrangian region corresponding to $\gamma$.
The displacement $\bfS$ is now spelled out using the {\it Zel'dovich\,}
approximation (Zel'dovich 1970),
$\bfS(\bfq,t)=-D(t)\, {\bf \nabla} \Phi(\bfq)$,
where $\Phi(\bfq)= \phi(\bfq,t)/ 4 \pi G \bar{\rho}(t) a^2(t) D(t)$
(with $G$ Newton's gravitational constant),
and $\phi(\bfq,t)$ is the gravitational potential.
Substituting in \equ{Llag} one obtains
\begin{equation}
\am (t)=-a^2(t) \dot{D}(t)\,\bar{\rho}_0 a^3_0 \,
\int_{\Gamma} (\bfq-\bar{\bfq}) \times {\bf \nabla} \Phi(\bfq) \,d^3q\, .
\label{eq:LZel1}
\end{equation}
The explicit growth rate is $L \propto a^2(t) \dot{D}(t)$, which is $\prop t$
as long as the universe is Einstein-de Sitter,
or close to flat (and matter dominated).

Next, assume that the potential is varying smoothly within the volume
$\Gamma$, such that it can be approximated by its {\it second-order} Taylor
expansion about the centre of mass,\footnote{
This is equivalent to assuming that the velocity field is well described
by its linear Taylor expansion, i.e., $v_i-\bar{v}_i\simeq\D_{ij}\,q'_j$.}
\begin{equation}
\Phi(\bfq') \simeq \Phi({\bf 0})+\left.\frac{\pa \Phi} {\pa q'_i}
\right| _{\bfq '=0} \,q'_i+
\frac{1}{2} \left.
\frac{\pa^2 \Phi}{\pa q'_i \pa q'_j} \right| _{\bfq'=0}
\,q'_i \, q'_j\;,
\label{eq:phiexp}
\end{equation}
where $\bfq'\equiv \bfq-\bar{\bfq}$.
Substituting in \equ{LZel1}, one obtains the basic TTT expression
for the $i$th Cartesian component:
\begin{equation}
L_i (t)= a^2(t) \dot{D}(t)\, \epsilon_{ijk}\, \D_{jl}\, I_{lk} \, ,
\label{eq:TTT}
\end{equation}
where $\epsilon_{ijk}$ is the antisymmetric tensor,
and the two key quantities are the {\it deformation\,} tensor at $\bfq'=0$,
\begin{equation}
\D_{ij}=-\left.\frac{\pa^2 \Phi}{\pa q'_i \pa q'_j} \right| _{\bfq'=0}\,,
\label{eq:deformation}
\end{equation}
and the {\it inertia\,} tensor of $\Gamma$,
\begin{equation}
I_{ij}=\bar{\rho}_0 a^3_0
\int_{\Gamma} q'_i \,q'_j \,d^3q' \, .
\label{eq:inertia}
\end{equation}

Note that only the traceless parts of the two tensors matter for the
cross product in \equ{TTT}. These are the velocity {\it shear\,}
or {\it tidal\,} tensor, $T_{ij}= \D_{ij}-(\D_{ii}/3) \delta_{ij}$,
and the traceless quadrupolar inertia tensor, $I_{ij}-(I_{ii}/3)\delta_{ij}$.
Thus, to the first non-vanishing order,
angular momentum is transferred to the proto-halo by the
gravitational coupling of the quadrupole moment of its mass distribution
with the tidal field exerted by neighboring density fluctuations.
The torque depends on the {\it misalignment\,} between the two.

\section{TTT in Simulations}
\label{sec:simu}

We summarize here the relevant issues concerning the simulation, the
halo finding, and the way we implement TTT. A more detailed description
is provided in Paper I.

The $N$-body simulation was performed as part of the
GIF project (\eg, Kauffmann \etal 1999) using the
adaptive P$^3$M code developed by the Virgo
consortium (Pearce \etal 1995; Pearce \& Couchman 1997).
As an example, we use a simulation of the $\tau$CDM scenario,
in which the cosmology is flat, with density parameter
$\omm=1$ and Hubble constant
$h=0.5$ ($H_0=100 \, h \kmsmpc$).
The power spectrum of initial density fluctuations is CDM with shape
parameter $\Gamma=0.21$, normalized to $\sigma_8=0.51$ today.
The simulation was performed in a periodic cubic box of side $84.55\hmpc$,
with $256^3$ particles of $1.0 \times 10^{10} h^{-1} M_\odot$ each.
Long-range gravitational forces were computed on a $512^3$ mesh, while
short-range interactions were calculated as in Efstathiou \& Eastwood (1981).
At late times ($z \lsim 3$), the gravitational potential
asymptotically matches a Plummer law with softening $\epsilon=36\hkpc$.
The simulation started at $z=50$ and ended at $z=0$.
The initial conditions were generated by displacing
particles according to the Zel'dovich approximation from an initial
stable `glass' state (e.g. White 1996).
More details are in Jenkins \etal (1998).

\begin{figure*}
\centerline{
\epsfxsize= 6.0 cm \epsfbox{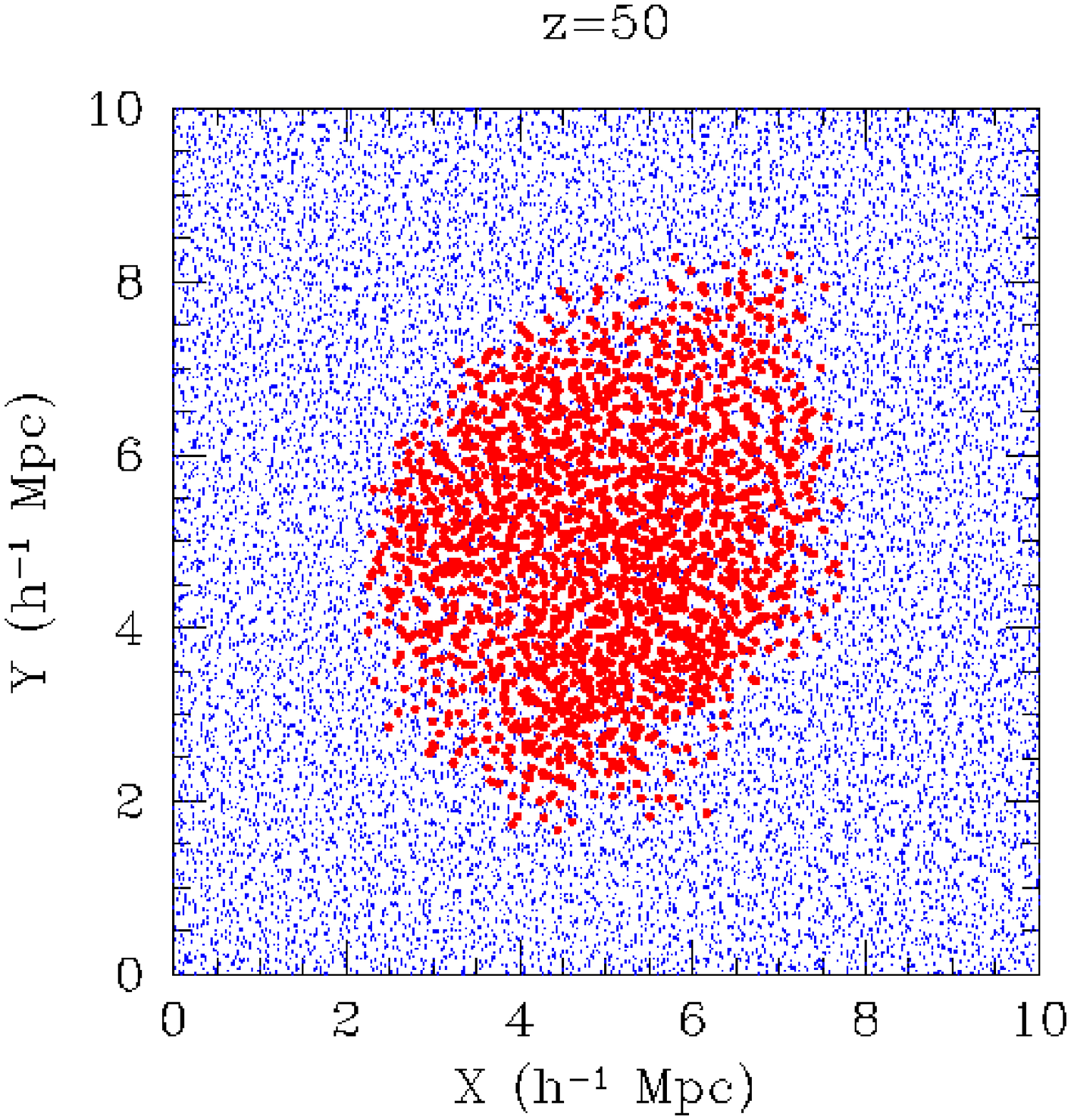}
\epsfxsize= 6.0 cm \epsfbox{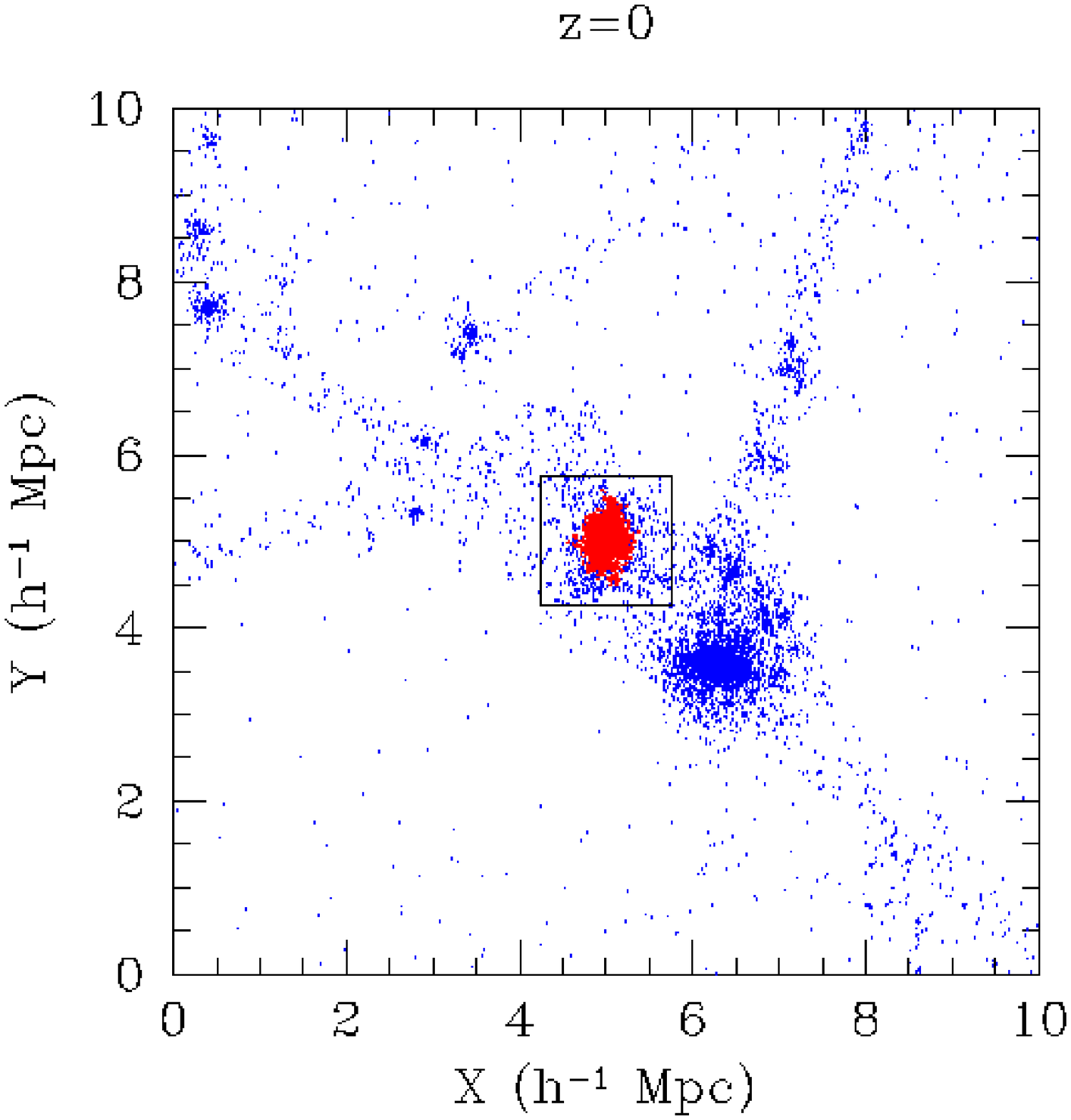}
\epsfxsize= 6.0 cm \epsfbox{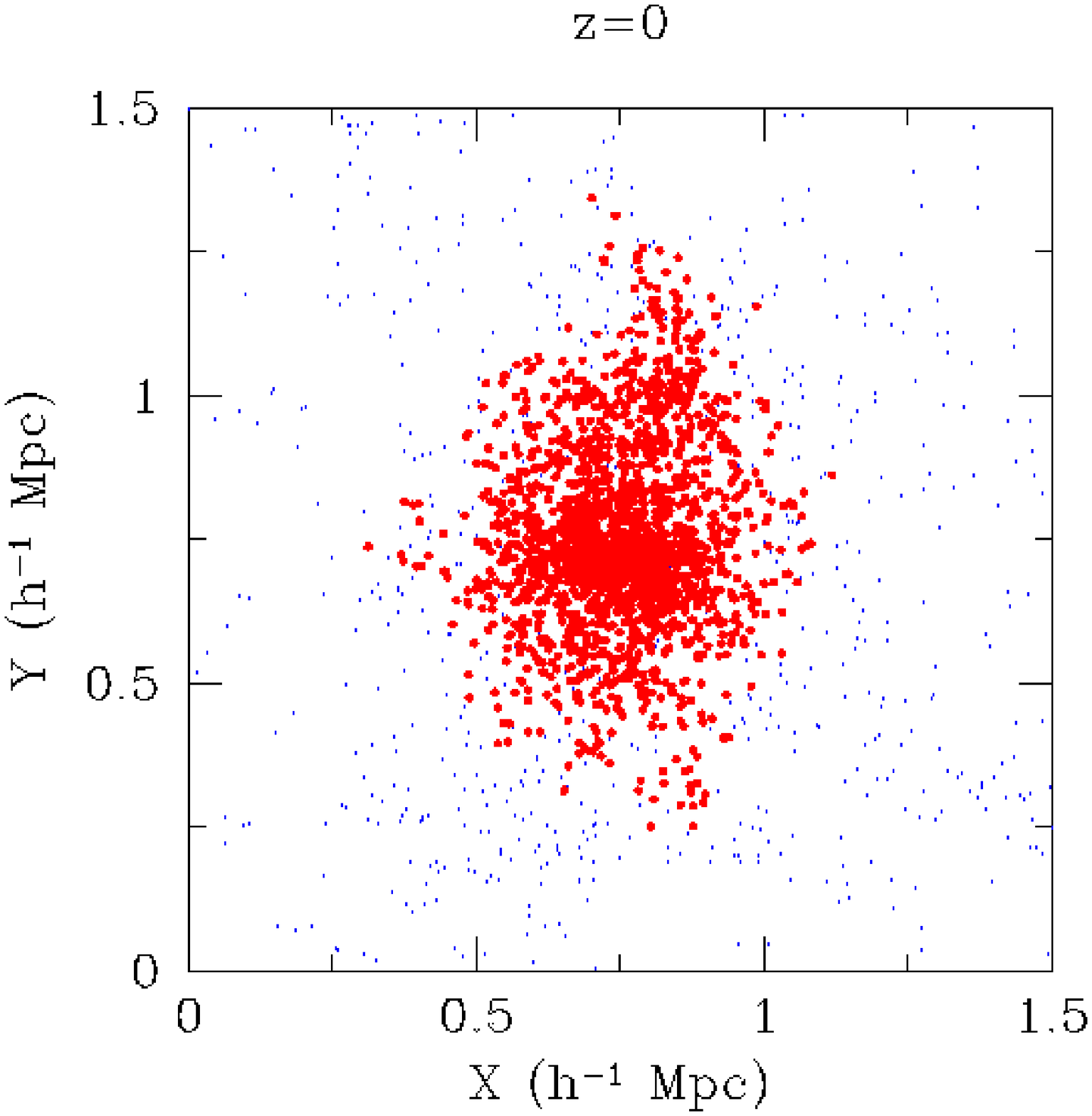}}
\caption{
An example of a halo at $z=0$ (center and right) and its proto-halo at $z=50$
(left),
embedded in the surrounding particles in a comoving slice about the halo
centre of mass.
The thickness of the slice is $5.4$ and $0.9 \,h^{-1}$ Mpc (comoving)
for the left, center and right panels respectively,
such that all the halo particles are included.
Particles belonging to the halo are marked
by filled circles, while points denote background particles.
Note that the halo is centred on one clump within a filament,
while the neighbouring haloes are not marked.
Note also that the proto-halo boundaries do not necesarily
follow the orientation of the filament.
Initial density contours in the region surrounding the proto-halo are
shown in Figure 4.
}
\label{fig:proto}
\end{figure*}

\fig{proto} shows a typical halo and its proto-halo, to help illustrating
the objects of our analysis.
We identify dark-matter haloes at $z=0$ using a standard friends-of-friends
algorithm with a linking length $0.2$ in units of the average inter-particle
distance.  This algorithm identifies regions bounded by a surface
of approximately constant density, corresponding to haloes with a mean
density contrast $\simeq 180$, in general agreement with spherical
perturbations whose outer shells have collapsed recently.
We then remove unbound particles from each halo, and consider only haloes which
contain more than 100 bound particles.
Our results turn out to be insensitive to the removal of unbound particles,
which is at a typical level of only a few per cent.

The robustness of our conclusions regarding TTT with respect to the
halo-finding algorithm (\eg, friends-of-friends versus fitting a spherical
or ellipsoidal density profile, Bullock \etal 2001a) is investigated
in another paper in preparation.
Note that all the haloes in our current sample are not subclumps of larger
host haloes, for which linear theory is not expected to be valid.
This excludes about 10 per cent of the haloes that are more massive
than $10^{12}\, h^{-1} \msun$ (Sigad \etal 2001).

The proto-halo regions $\Gamma$ are defined by simply tracing all the
virialized halo particles, as identified today, into their Lagrangian
positions.  In most cases, most of the proto-halo
mass is contained within a
simply connected Lagrangian region, but in some cases ($\sim 15$ per cent)
the proto-halo may be divided into
two or three
compact regions which are connected by thin filaments.
Nearly 10 per cent of the proto-haloes are characterized by extended
filaments departing from a compact core.
A typical case is illustrated in \fig{proto}, where we show a halo
at $z=0$ and its proto-halo at $z=50$, embedded in the surrounding particles
in a comoving slice that includes most of the halo particles.

For each proto-halo, we compute the Lagrangian inertia tensor
by direct summation over its $N$ particles of mass $m$ each:
\begin{equation}
I_{ij}=m \sum_{n=1}^{N} q_i^{'(n)} q_j^{'(n)}\;,
\end{equation}
with $\bfq^{'(n)}$ the position of the $n$-th particle with respect
to the halo centre of mass.

The shear tensor at the proto-halo centre of mass is computed by
first smoothing the potential used to generate the initial Zel'dovich
displacements, and then differentiating it twice with respect to the
spatial coordinates ({\it Method 1}).
Smoothing is done using a top-hat window function, while derivatives are
computed on a grid. The top-hat smoothing radius, in comoving units,
is taken to be defined by the halo mass via $(4 \pi/3) \bar\rho_{0} R^3=M$.
In Paper I we also tested two alternative methods for computing the shear
tensor: {\it Method 2},
in which the smoothing has been replaced by minimal variance fitting,
and {\it Method 3}, also with minimal variance fitting, but
in which we considered only the shear generated by the
density perturbations lying outside the proto-halo volume.
The TTT predictions for the halo spin based on the three methods
were found to be of similar quality at $z=0$, while
Methods 2 and 3 are slightly more accurate at very high redshift.
We adopt method 1 here because it does not depend explicitly on the detailed
shape of the proto-haloes, which prevents spurious correlations between
$T$ and $I$ due to the way $T$ is computed.  Also, this is the only method
applicable in analytic and semi-analytic modelling.

\section{Alignment of Inertia and Shear}
\label{sec:T-I}

\begin{figure*}
\epsfxsize= 12 cm \epsfbox{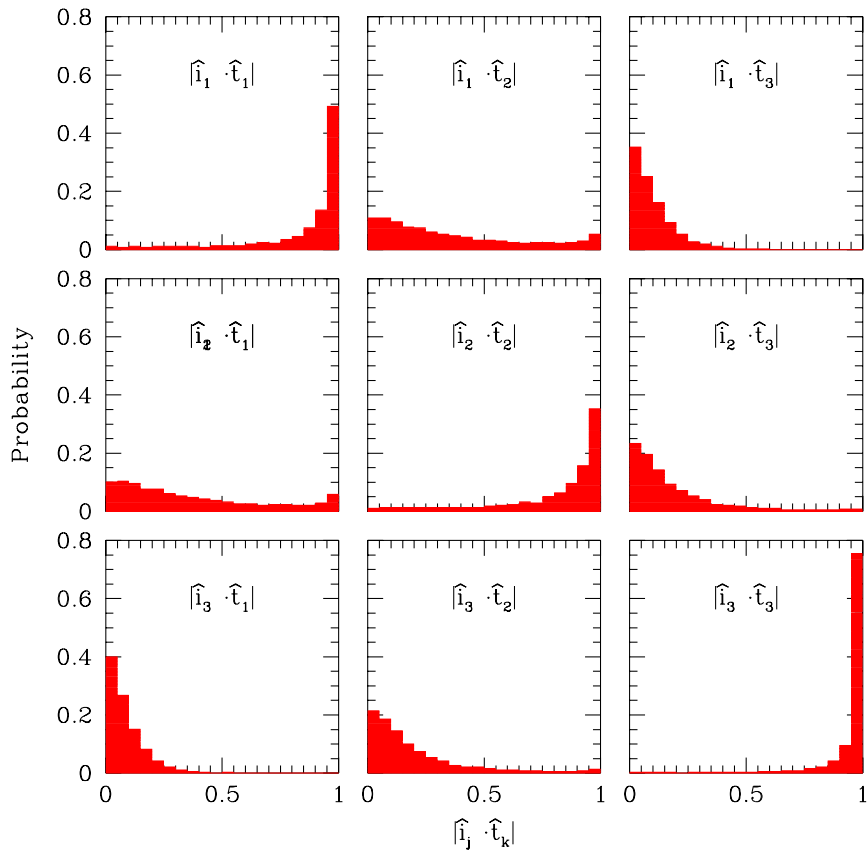}
\caption{
Alignments of $I$ and $T$.
Distributions of the cosine of the angle between the
directions of the principal axes of the inertia tensor ($\hat{\bfi}_j$)
and of the velocity shear tensor ($\hat {\bft}_k$).
The corresponding eigenvalues are ordered $i_1\geq i_2\geq i_3$
and $t_1\leq t_2\leq t_3$, such that $1$ denotes the major axis and $3$
denotes the minor axis.
The first two moments of the distributions are given in
\equ{mom1} and \equ{mom2}.
}
\label{fig:t-i}
\end{figure*}

In Paper I, we evaluated the success of TTT in predicting
the final halo angular momentum for a given proto-halo embedded in a given
tidal environment at the initial conditions. For a deeper understanding of how
TTT works, we now proceed to investigate the key players:
the inertia and shear tensors and in particular the cross-talk between them.
Note, in \equ{TTT}, that angular momentum is generated only due to
{\it misalignments} of the principal axes of these tensors.
It has been assumed, in many occasions, that these two tensors are
largely uncorrelated, the inertia tensor being a local property while
the shear tensor is dominated by large-scale structure external to the
proto-halo (e.g. Hoffman 1986a, 1988a; Heavens \& Peacock 1988;
Steinmetz \& Bartelmann 1995; Catelan \& Theuns 1996).
Such a lack of correlation would have led
to relatively large spins and would have provided a specific statistical
framework for TTT.  On the other hand, a strong correlation between
$I$ and $T$ would have led to relatively small spins,
and would have invalidated some of the predictions based on
the assumption of independence of $I$ and $T$.
Moreover, a correlation of this sort could provide the missing clue for
the special characteristics of the Lagrangian proto-halo regions $\Gamma$,
those that will eventually evolve into virialized haloes.

We address the correlation between the principal axes of $I$ and
$T$ at the proto-halo centres of mass
by first showing in \fig{t-i} the probability distributions of the cosines of
the angles between them. The eigenvectors $\hat {\bf i}_j$ and
$\hat{\bf t}_k$ are labeled in such a way
that the corresponding eigenvalues are ranked $i_1\geq i_2 \geq i_3$ and
$t_1\leq t_2 \leq t_3$, namely the major axes are denoted by 1 and
the minor axes by 3 (and note, for example, that $t_1$ is the direction
of maximum compression).\footnote{The principal axes of $T_{ij}$ coincide
with those of ${\cal D}_{ij}$, because the difference of the two tensors is
a scalar matrix.}
The highest spike (near a cosine of unity) indicates a very strong
alignment between the minor axes of the two tensors, and the second-highest
spike refers to a strong alignment between the major axes.
The alignment between the intermediate axes is also apparent but somewhat
weaker.
The peaks near a cosine of zero indicate a significant tendency for
orthogonality between the major axis of one tensor and the minor axis
of the other.
There is a somewhat weaker orthogonality between the intermediate and minor
axes, and the weakest orthogonality is between the major and intermediate
axes.

\begin{figure*}
\centerline{
\epsfig{figure=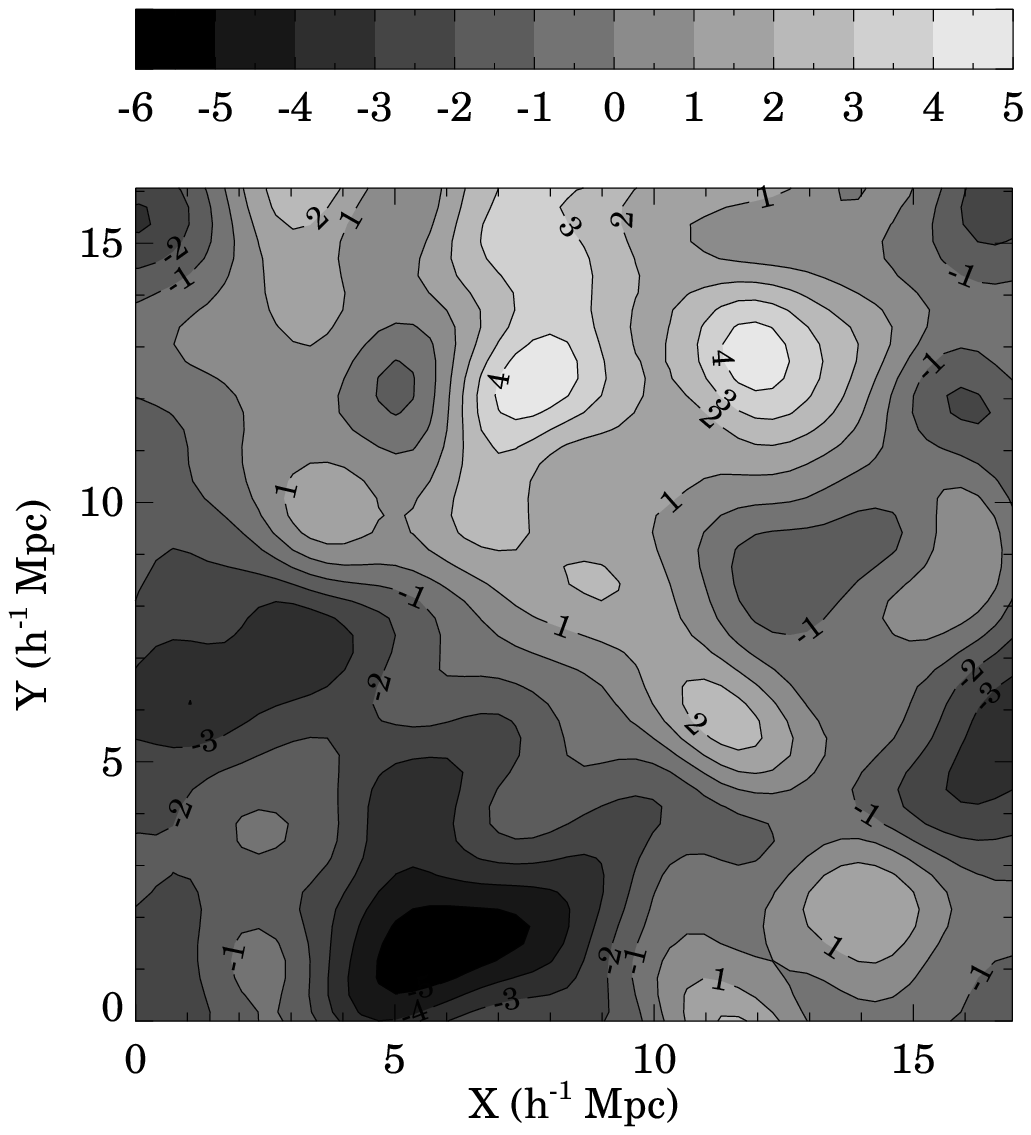,width=10cm}
\hspace {-1.9cm}
\epsfig{figure=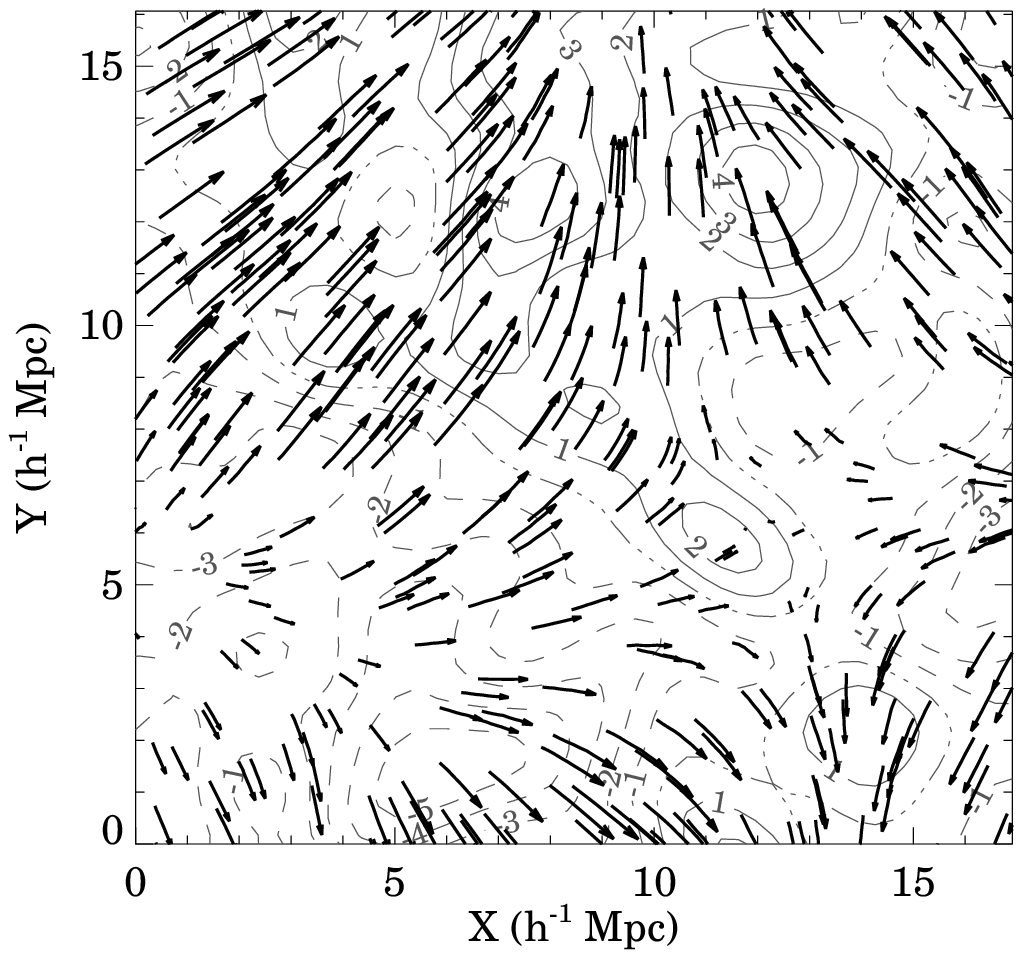,width=10cm}}
\vspace{-1.cm}
\centerline{
\epsfig{figure=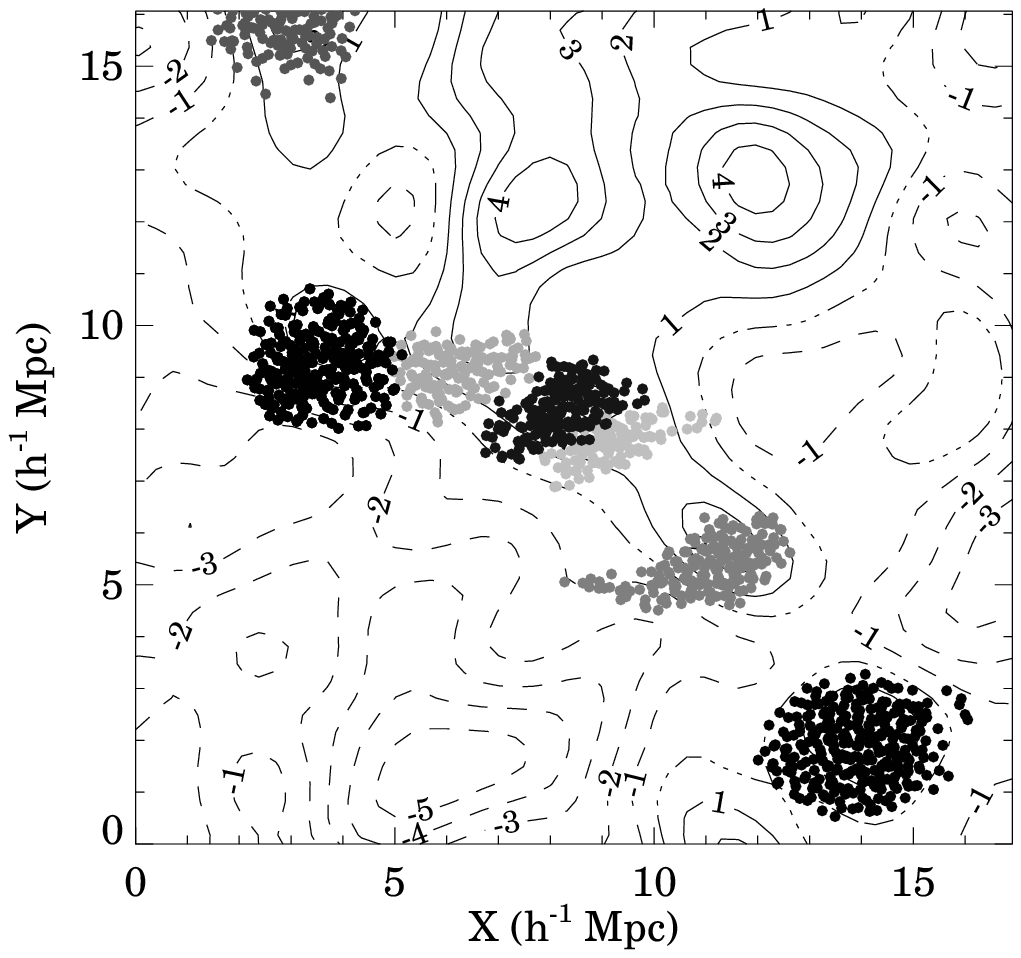,width=10cm}
\hspace {-1.75cm}
\epsfig{figure=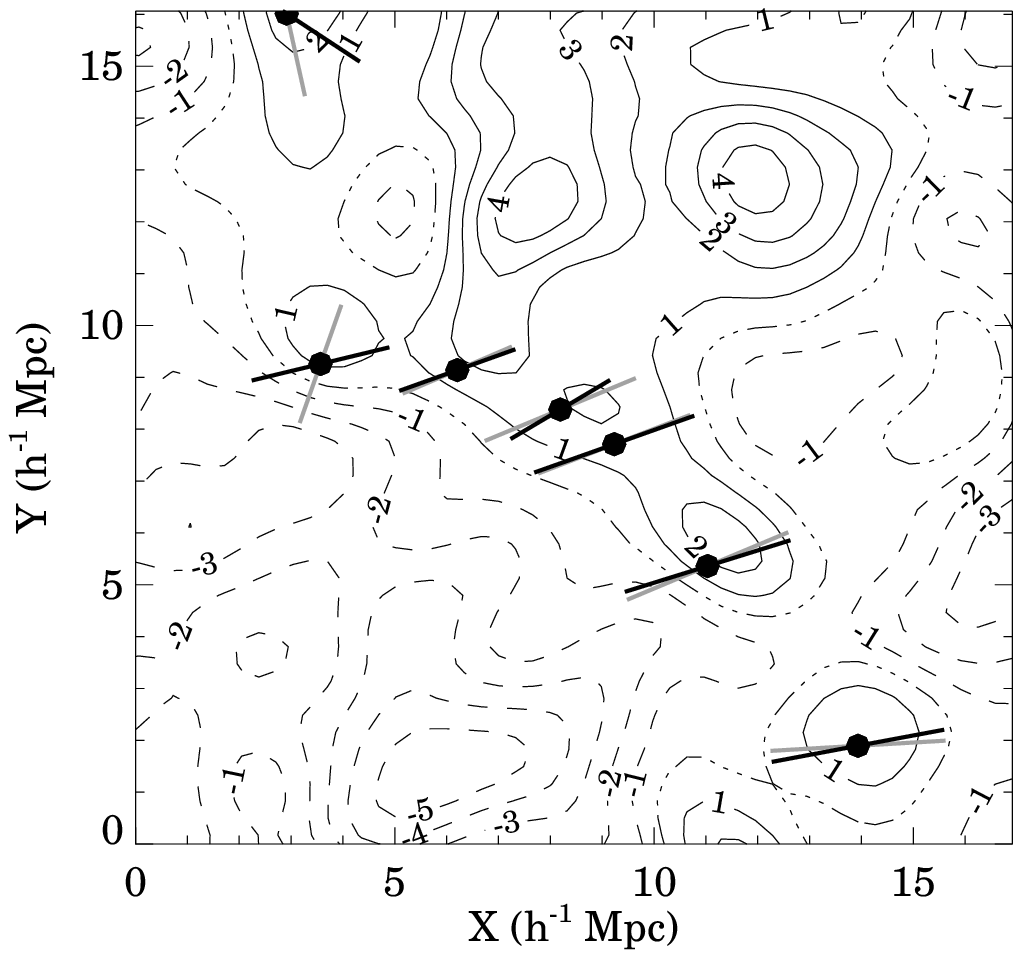,width=10cm}}
\caption{
Examples of the correlation between I and T.
The {\it top panels\,} show maps of the density (left) and
velocity (right) fields, at $z=50$, in a section of an $X-Y$ plane
from the simulation box.
The fields are smoothed with a top-hat window of radius $0.95 \hmpc$
corresponding to a 100-particle halo.  The contours
refer to the density contrast linearly extrapolated to $z=0$.
The {\it bottom panels\,} show all the proto-haloes whose centres of mass
lie within one smoothing length of the plane
(A cluster-size halo associated with the high density peaks at the top
is not shown because its centre lies further away from the plane).
The left panel shows the
projection of the proto-halo particle positions.  The haloes contain,
from left to right, 179, 257, 143, 168, 100, 183, and 300 particles.
The right panel shows the projections of the major axes of $I$ (dark lines)
and $T$ (light lines) about the centres of mass (filled circles).
The line length is proportional to the projection of a unit
vector along the major axis on to the $X-Y$ plane.
To set the scale, we note, for example, that
the principal axis of the inertia tensor of the proto-halo at the centre
of the panel lies almost exactly in the plane shown.
}
\label{fig:ex1}
\end{figure*}

\begin{figure*}
\centerline{
\epsfig{figure=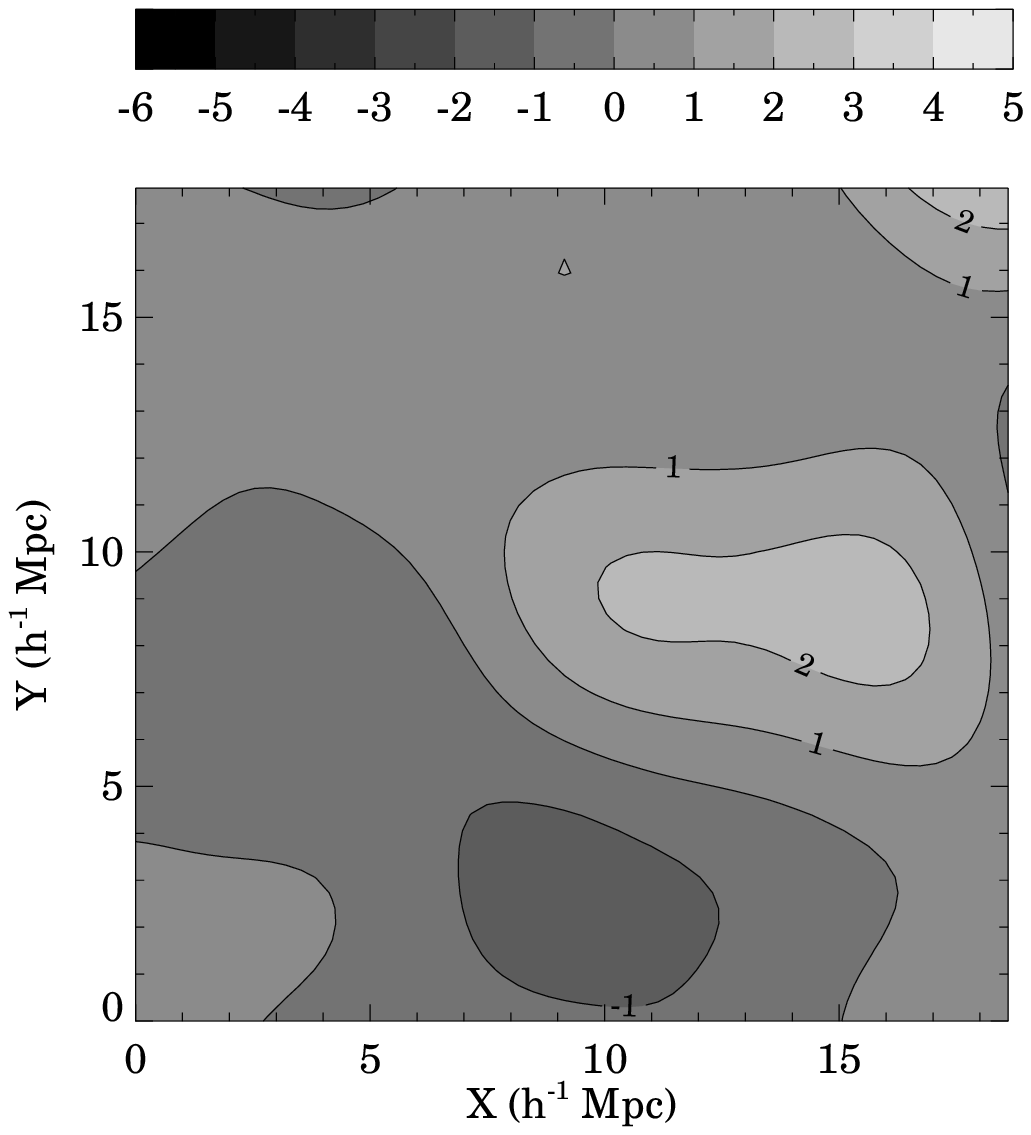,width=10cm}\hspace {-1.9cm}
\epsfig{figure=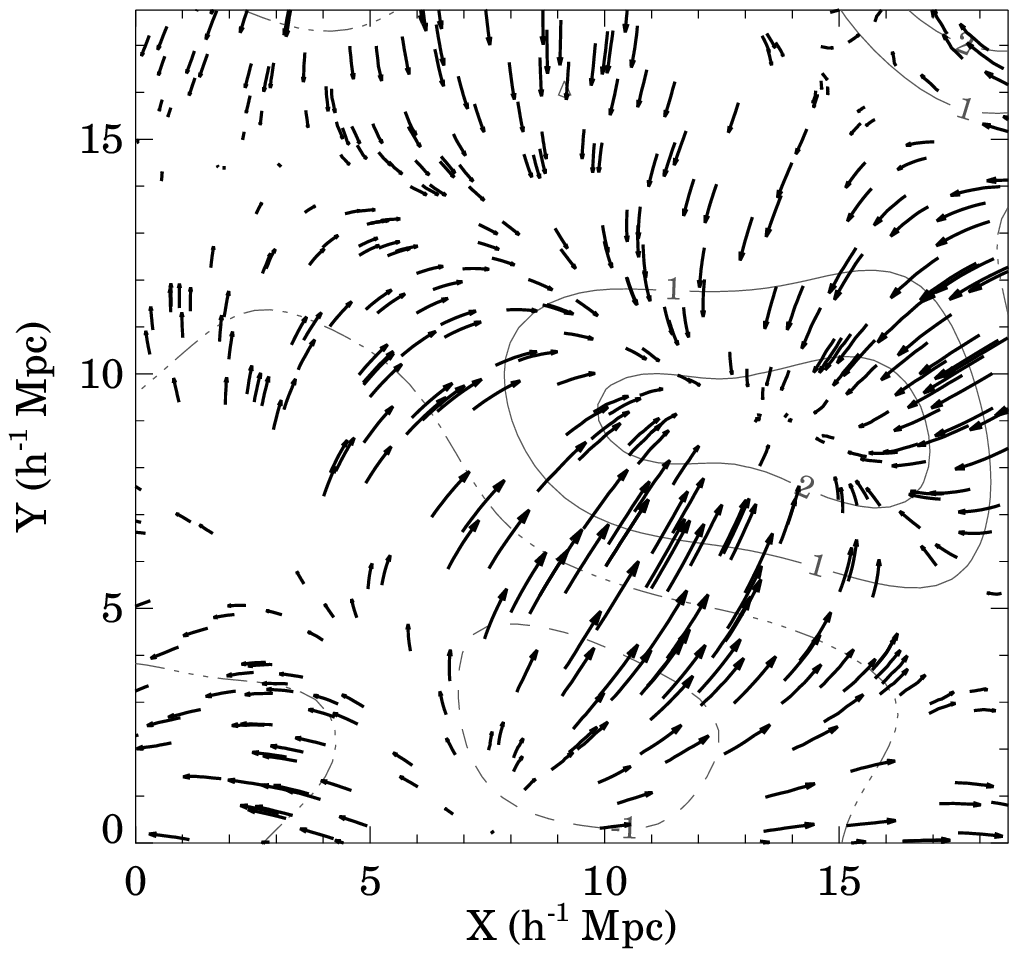,width=10cm}}
\vspace{-1.cm}
\centerline{
\epsfig{figure=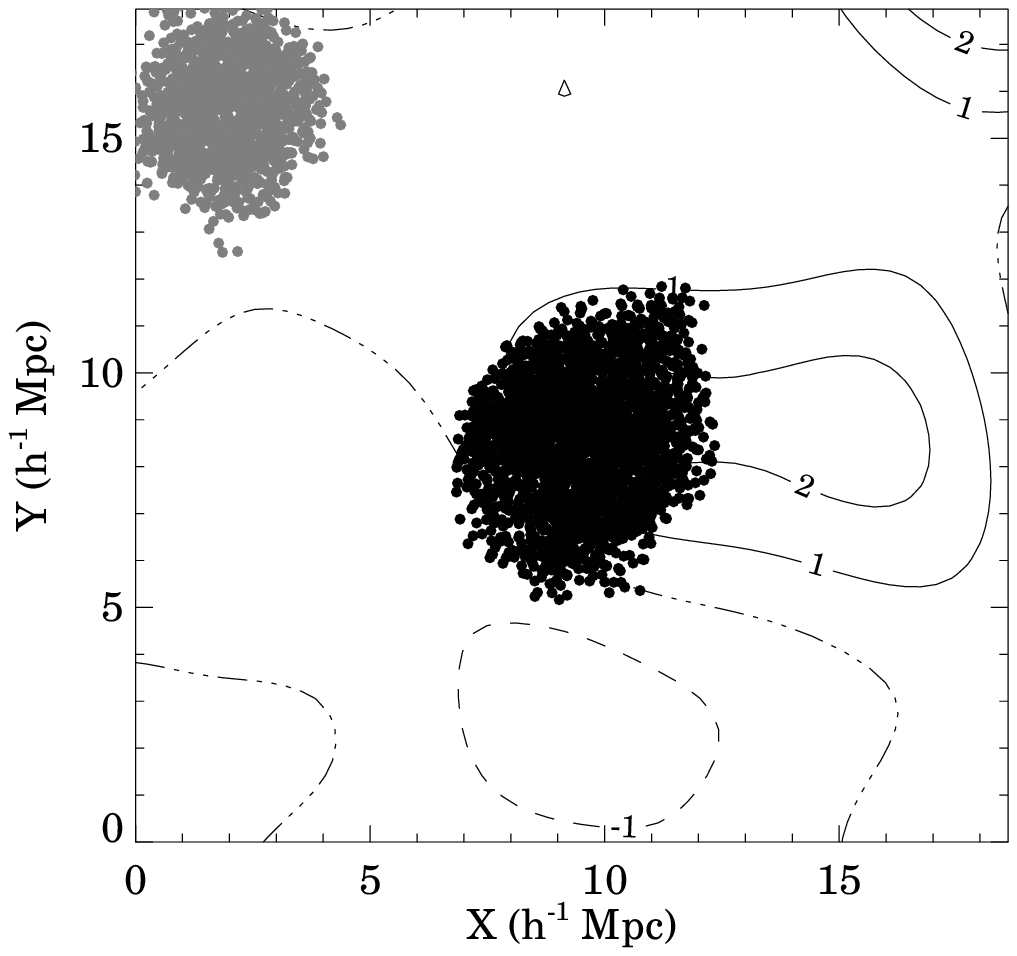,width=10cm}\hspace {-1.75cm}
\epsfig{figure=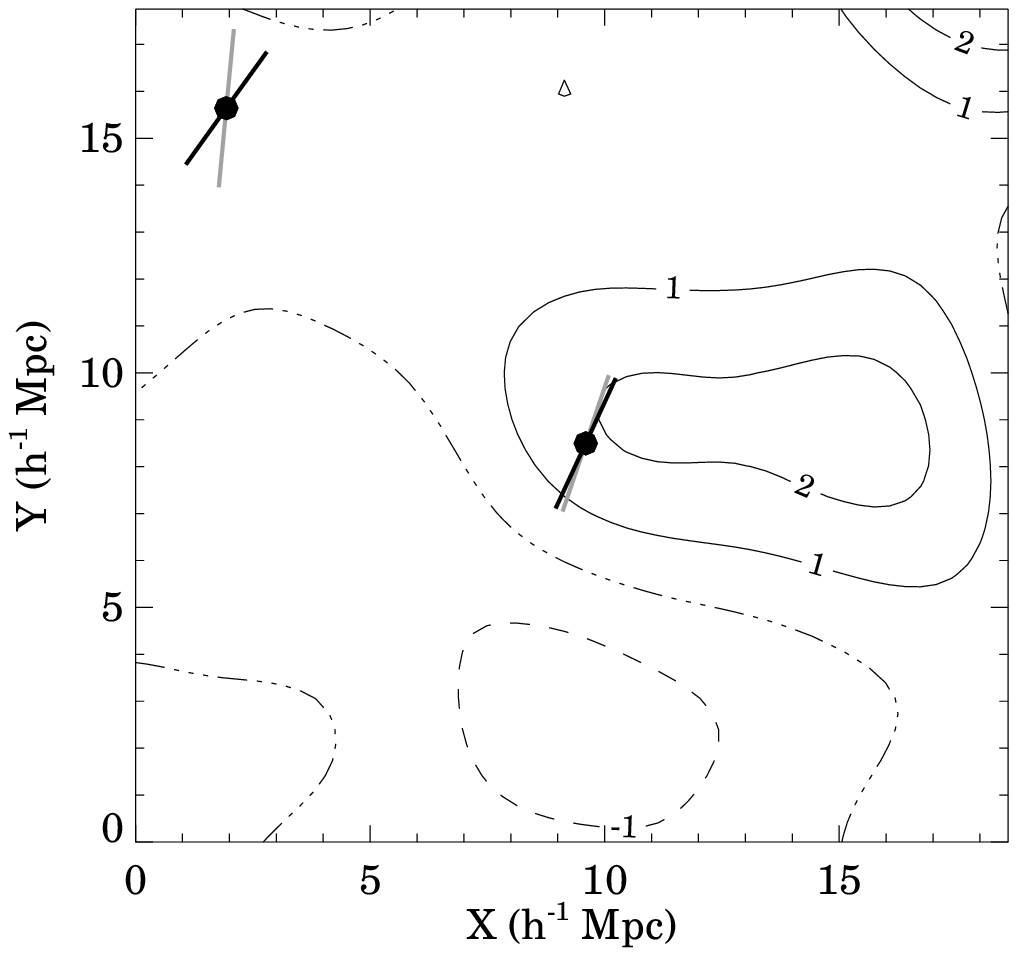,width=10cm}}
\caption{Same as \fig{ex1}, but for a different region of the simulation,
and showing only the haloes of more than 1000 particles.
The haloes contain, from left to right, 1065 and
2003 particles.  The smoothing length of $2.58 \hmpc$ now corresponds to a
2000-particle proto-halo.
}
\label{fig:ex2}
\end{figure*}
The first two moments of these distributions are
\begin{equation}
\mu_{ij}=\langle | \hat{\bfi}_i \cdot \hat{\bf t}_j |
\rangle=\begin{pmatrix}
0.838 & 0.363 & 0.106  \\
0.370 & 0.793 & 0.196  \\
0.091 & 0.206 & 0.935
\end{pmatrix}\;,
\label{eq:mom1}
\end{equation}
and
\begin{equation}
\nu_{ij}=\langle (\hat{\bfi}_i \cdot \hat{\bf t}_j )^2\rangle=
\begin{pmatrix}
0.759 & 0.219 & 0.023  \\
0.225 & 0.692 & 0.083  \\
0.017 & 0.089 & 0.895
\end{pmatrix}\;.
\label{eq:mom2}
\end{equation}
Based on the higher-order moments and the
number of haloes in the sample, the statistical uncertainty
in these mean quantities
is between 0.001 and 0.003, depending on the matrix element considered.
In order to estimate the systematic errors due to the finite number
of particles in each halo, we re-measured
the moments considering only the haloes containing 1000 particles or more.
The results agree with those dominated by smaller haloes of $\geq 100$
particles at the level of a few per cent. For example, we obtain
\begin{equation}
\mu_{ij}=\begin{pmatrix}
0.792 & 0.418 & 0.142  \\
0.433 & 0.733 & 0.226  \\
0.117 & 0.243 & 0.907
\end{pmatrix}\;,
\end{equation}
with a typical statistical uncertainty of $0.010$.

The values of the matrix elements $\mu_{ij}$ and $\nu_{ij}$ can be used
to quantify the degree of correlation between the shear and inertia tensors.
Note that a perfect correlation corresponds to a diagonal, unit matrix
both for $\mu$ and $\nu$,
while perfect independence corresponds to all the elements being equal
at $1/2$ (for $\mu$) or $1/3$ (for $\nu$).
Note that, by definition, $\sum_j \nu_{ij}=\sum_i \nu_{ij}=1$.
To characterize the above matrices with single numbers,
we define a correlation parameter for $\mu_{ij}$ by
\begin{equation}
c_\mu=\f{1}{9} \sum_{i,j=1}^3 \f{\mu_{ij}-1/2}{\delta_{ij}-1/2}=
\f{1}{3}+\f{2}{9} ( \mu_{ii}-\sum_{i\ne j}\mu_{ij} )\; ;
\end{equation}
it obtains the values zero and unity for minimum and maximum
correlation respectively.
Similarly, we define a correlation parameter using $\nu_{ij}$,
\begin{equation}
c_\nu=\f{1}{9} \sum_{i,j=1}^3 \f{\nu_{ij}-1/3}{\delta_{ij}-1/3}=
\f{1}{2}\left(\nu_{ii}-1 \right)\;.
\end{equation}
We find in the simulation $c_\mu=0.61$ and $c_\nu=0.67$, indicating a strong
correlation between $I$ and $T$.

We have repeated the above correlation analysis for the two alternative methods
of computing the shear tensor described in \S 4 of Paper I.
When we minimize smoothing and include only the part of the shear tensor
that is generated by fluctuations external to the proto-halo ({\it Method 3}),
the correlations are very similar to those obtained with top-hat smoothing,
indicating that these are indeed the
correlations of physical significance that we are after.
When we minimize smoothing and include the contribution from internal
fluctuations ({\it Method 2}), the correlations still show but they get
significantly weaker because of the additional noise in the computation of $T$.
Thus, our standard way of computing $T$ ({\it Method 1}) seems to pick up
properly the correlation with $I$, and we proceed with it.

Note that $T$-$I$ correlations 
are unavoidable in the framework of the Zel'dovich
approximation, where a halo consists by the matter that has shell-crossed.
However, the strength of the expected correlations has still to be worked out.

The implication of the $T$-$I$ alignment is that
proto-halo spins are due to small residuals from this correlation.
This means that the haloes acquire significantly less angular momentum than
one would have expected based on a simple dimensional analysis that
ignores the correlation.
We quantify the effect by a comparison to an artificial case where the
alignments are erased.
For each halo in our sample, we randomize the relative directions of the
eigenvectors of the inertia and tidal tensors (by a three-dimensional
rotation with random Euler angles),
and then re-compute the angular momentum using \equ{TTT}.
A vast majority of the haloes, about 88 per cent, are found to be associated
with a larger spin after the randomization procedure.
The $T$--$I$ correlation is found to reduce the halo spin
amplitude with respect to the randomized case by an average factor of
$3.1$, with a comparable scatter of $3.2$.
This heads towards explaining why haloes have such
low values of the dimensionless spin parameter, $\lambda \sim 0.035$
on average (Barnes \& Efstathiou 1987; Bullock \etal 2001b),
namely a very small rotational energy compared to gravitational
or kinetic energy.

\section{Proto-halo Regions}
\label{sec:gamma}

The strong correlation between the principal axes of the
inertia and tidal tensors promises very interesting implications
on galaxy-formation theory. It indicates that the tidal
field
plays a key role in determining the locations and shapes of proto-haloes,
and therefore may provide a useful tool for identifying proto-haloes in
cosmological initial conditions.

As a first step, we try to gain intuitive understanding of the nature
of the correlation between $T$ and $I$ by inspecting a few proto-haloes and
their cross-talk with their cosmological environment.
\fig{ex1} and \fig{ex2} show a few examples, in regions of the simulation box
containing several proto-haloes.
In the top panels we show maps of the density and velocity fields
at the initial conditions, to indicate some of the qualitative properties
of the shear field.
We focus on a section of one plane at at time, and project the velocity
field onto it.
The fields are smoothed with a top-hat window corresponding to the typical
halo masses
shown
in that region: in \fig{ex1}, the smoothing radius is $0.95 \hmpc$,
to match the haloes that contain 100 to 300 particles each, while
in \fig{ex2} it is $2.58 \hmpc$, to match the haloes of 1000 to 2000
particles.
The smoothed density contrast is linearly extrapolated to $z=0$.

We then introduce, in the bottom panels, the proto-haloes and their inertia
tensors.
We present all the proto-haloes whose centres of mass lie within one smoothing
length of the plane by showing, in the bottom-left panels,
the projection of the proto-halo particle positions onto the plane.
We see a tendency for an association between the proto-halo centres of mass
and density peaks smoothed on a halo scale, but this association seems
to be far from perfect (e.g. near the centre of the frame in \fig{ex1}).
Note that the single plane shown does not allow an accurate identification
of the peak location, and therefore a definite evaluation of the association
of proto-haloes and peaks cannot be properly addressed by observing these maps.
In order to quantify the correlation between proto-haloes and peaks,
we identified peaks in the linear density field, smoothed with a top-hat
window that contains $n$ (with $n$=100, 1000 and 10000) particles,
and determined the nearest peak to the centre of mass of each of the
proto-haloes that contain
$n$ ($\pm 10$ per cent)
particles.
Independently of the halo mass, we find that only $\sim 35-45$
per cent of the proto-halo centers lie within one
smoothing radius from the nearest peak, and $\sim 60-65$
per cent lie within two smoothing radii.
We conclude that while the proto-haloes tend to lie in high density regions,
their centres do not coincide with the local density maxima, and in fact the
spatial
correlation between the two is quite weak.

In the bottom-right panels, we stress the projections of the major
axes of the inertia and shear tensors about the centres of mass,
and indicate by the length of the line the cosine of the angle between the
axis and the plane shown.
Note, for example, that the major
axes of those haloes in \fig{ex1} that reside near the centre of the
frame and towards the bottom right happen to lie almost perfectly in the
plane shown.
All the proto-haloes that show a significant deviation from spherical symmetry
have their major inertia axis strongly aligned with the first principal
shear axis at their centre of mass.
This alignment reflects the fact that the largest compression flow
towards the centre of the proto-halo is along the major inertia axis of the
proto-halo. Indeed, this is exactly what is required
in order to compress the elongated proto-halo into the more centrally
concentrated and quite spherical
configuration identified by the halo finder after collapse and virialization.

The detected alignment is a clear manifestation of the crucial role played
by the external shear in determining the shape of the Lagrangian volume of
the proto-halo.  One way to interpret this alignment is as follows.
The compression along the major axis of $T$ is associated with a flow of matter
from the vicinity into the proto-halo. This matter comes from relatively large
distances and it is therefore responsible for a large inertia moment along
this axis in the proto-halo configuration.  In contrast, the dilation along
the minor axis of $T$ causes matter to be tidally stripped, thus leading to
a relatively small inertia moment of the proto-halo along this axis.
This picture, in which the boundaries of the proto-halo are fixed by the
push and pull of the external mass distribution, is in some sense
the opposite of the common wisdom, where the crucial factor is assumed to
be the self-gravity attraction. This other approach, which predicts that
collapse first occurs along the minor inertia axis, is based on the simplified
model of the collapse of an isolated ellipsoidal perturbation that starts
at rest or comoving with the Hubble flow
(Lin, Mestel \& Shu 1965; Zel'dovich 1965).

\begin{figure*}
\centerline{\epsfxsize=12 cm \epsfbox{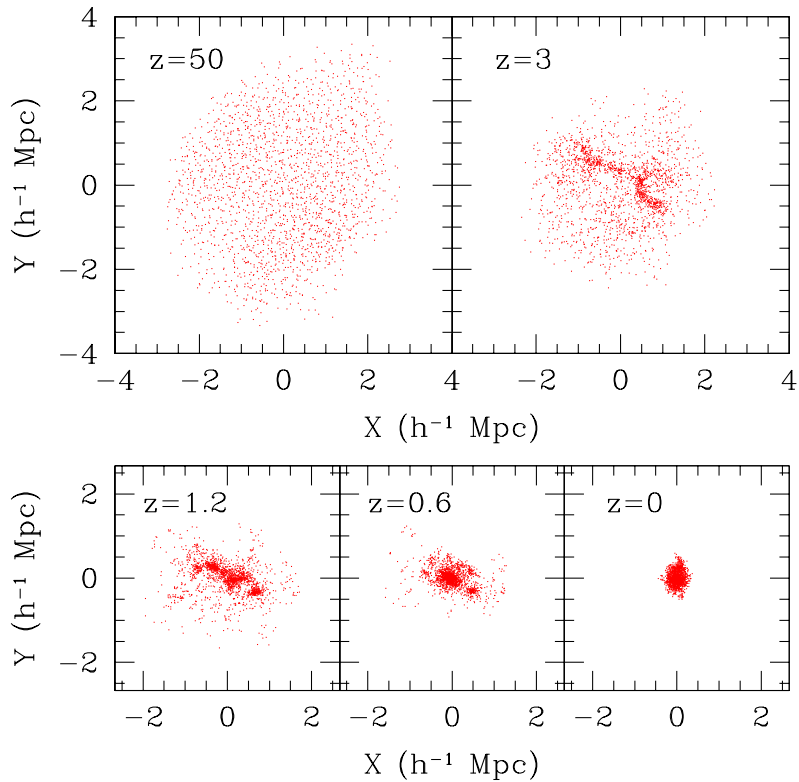}}
\caption{Time evolution of a protohalo.
Shown are the projected positions of the halo particles at different redshifts
in a fixed comoving box about the centre of mass.
The first collapse along the major axis of inertia (and the major axis of
the tidal field) leads to the formation of a transient filament, which
breaks into sub-clumps. These sub-clumps then flow along the filament
and merge into the final halo}
\label{fig:hist}
\end{figure*}

Snapshots of the evolution of a typical halo
(the same massive halo shown in \fig{proto} and \fig{ex2})
are shown in \fig{hist}.
Plotted are the positions at different epochs
of the particles that form the halo identified at $z=0$.
The proto-halo first collapses along its major and intermediate
inertia axes, giving rise to an elongated structure made of sub-clumps.
The final halo is then assembled by merging and accretion along
the axis of the elongated filament.
This transient sub-halo filament lies along the large-scale filament
in which the final halo is embedded (see \fig{proto} and \fig{ex2}).
It indicates that the late stages of halo formation are associated
with flows along preferential directions aligned with the cosmic web
(see also Colberg \etal 1999).

It has often been assumed that the boundaries of proto-haloes could be
identified with some threshold iso-density contours about density maxima.
(Bardeen \etal 1986; Hoffman 1986b, 1988b;
Heavens \& Peacock 1988; Catelan \& Theuns 1996).
In this case, the density distribution in a proto-halo could be approximated
by a second-order Taylor expansion of the density profile about the peak.
Figures \ref{fig:ex1} and \ref{fig:ex2}
demonstrate that this assumption is going to fail in most
cases shown (except, perhaps, the halo at the bottom-right of \fig{ex1}).
In a typical proto-halo, the boundaries are not determined by the self gravity
due to the local mass distribution within the proto-halo but rather by the
tidal field due to the external mass distribution.
Such a proto-halo is typically embedded within a large-scale elongated density
ridge, but is not necessarily centred on a local density peak. The
compressions exerted by the void regions surrounding the ridge define
the compression axes of the shear tensor in directions perpendicular
to the ridge, while the tides from the other parts of the ridge cause
large-scale 
dilation along the ridge.  The associated push and pull of mass along
these directions make the large inertia axes of the proto-halo lie
perpendicular to the ridge, and the minor inertia axis lie parallel to the
ridge.  This understanding should be translated to a practical recipe for
identifying the boundaries of proto-haloes.
At later stages we also see an internal compression and merging along 
the filament, namely the minor inertia axis of the proto-halo.

\section{Other Properties of Proto-haloes}
\label{sec:properties}

We continue with a complementary investigation of the properties of
proto-haloes, using several different statistics that may, in particular,
distinguish them from random Gaussian peaks.

\begin{figure}
\epsfxsize= 8 cm \epsfbox{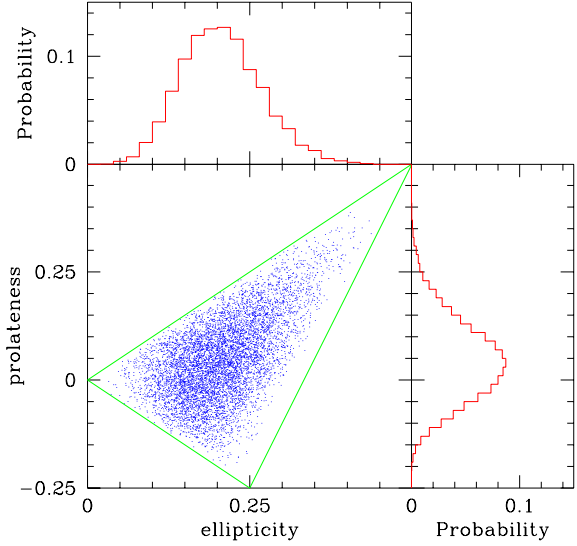}
\caption{
Prolateness and ellipticity for the inertia tensor of proto-haloes.
The average and
standard deviation values are
$\la e\ra=0.206$ and $\sigma_e=0.062$
and
$\la p \ra=0.048$ and $\sigma_p=0.092$.
}
\label{fig:I_ep}
\end{figure}

First of all, we study the shapes of the proto-haloes in Lagrangian space
using their inertia tensors.
A complete description of the statistical properties of the eigenvalues
of the inertia tensor (a symmetric, positive definite matrix)
can be obtained in terms of the three following parameters:
the trace $\tau=i_1+i_2+i_3$, the ellipticity $e=(i_1-i_3)/2\tau$,
and the prolateness $p=(i_1 -2i_2 + i_3)/ 2\tau$.
A perfect sphere has $e=p=0$. A flat circular disc (ultimate oblateness)
has $e=1/4$ and $p=-1/4$. A thin filament (ultimate prolateness)
has $e=1/2$ and $p=1/2$. Thus, $e$ measures the deviation from sphericity,
and $p$ measures the prolateness versus oblateness.
In \fig{I_ep} we show the joint distribution of ellipticity and
prolateness, and the probability distribution marginalized along each
axis, for our proto-haloes at $z=50$.
The boundaries $e \geq -p$, $e \geq p$, and $p\geq 3e-1$
arise from the conditions $i_1 \geq i_2$, $i_2 \geq i_3$, and
$i_3 \geq 0$, respectively.
As a consequence, the data-points populate only a triangle
in the $e-p$ plane, with vertices at $(0,0)$, $(1/2,1/2)$ and $(1/4,-1/4)$.
This introduces a correlation between $e$ and $p$ at high ellipticities
($e>1/4$).
We find that only 3 per cent of the proto-haloes are nearly spherical,
with $e<0.1$ (or $|p|<0.1$).
Most of the proto-haloes, 74 per cent,
have moderate but significant ellipticities
in the range $0.1< e<0.25$. A significant fraction, 23 per cent, have extreme
ellipticities of $e>0.25$ (and almost all are prolate configurations).
About 68 per cent of the proto-haloes are prolate, $p>0$,
and about 2.3 per cent are extremely prolate, $p>0.25$.
The average and standard deviation for $e$ and $p$ are
$0.206 \pm 0.062$ and $0.048\pm 0.092$, respectively.
It is pretty surprising to note that, despite the pronounced difference of
the proto-halo boundaries from iso-density contours about density peaks,
the shape distribution of the two are
not very different.
Bardeen \etal (1986, section VII) quote for overdensity patches about
$2.7\sigma$ peaks: $0.17 \pm 0.07$ and $0.005 \pm 0.098$ for $e$ and $p$
respectively. Using their equations 7.7 and 6.17 for $1\sigma$ peaks,
which we found to be more appropriate for proto-haloes,
we get $0.19 \pm 0.08$ and $0.04\pm 0.11$ for $e$ and $p$ respectively.

\begin{figure}
\epsfxsize=8cm \epsfbox{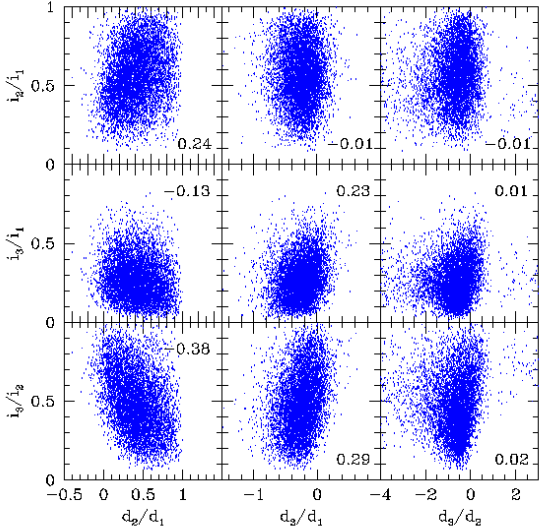}
\caption{Ratios of the eigenvalues of the inertia tensor versus
ratios of the eigenvalues of the deformation tensor (evaluated at the
halo centre of mass) for the entire halo population in the simulation.
Also shown is the corresponding linear correlation coefficient for
the joint distribution in each panel.}
\label{fig:De_Ie}
\end{figure}

In \se{T-I} we found that the principal axes of the inertia and
tidal (and deformation) tensors tend to be aligned,
which means that proto-haloes will have their dominant collapse along
their major axes of inertia.
We now test whether the shape of the proto-halo
is correlated with the relative strength of the eigenvalues of the
deformation tensor, i.e., the compression factors
along the principal axes of ${\cal D}$.\footnote{
We deal here with ${\cal D}$ rather than $T$ because we expect the self
gravity of the proto-halo (described by the diagonal terms of ${\cal D}$)
to also play a role in shaping it,
as in the basic formulation of the Zel'dovich approximation.}
In \fig{De_Ie}, we show the joint distributions of the ratios $i_j/i_k$ and
$d_j/d_k$ between the eigenvalues of the inertia and deformation tensors
($d_i=t_i+{\cal D}_{ii}/3$).\footnote{
We prefer here these ratios over the $e$ and $p$ parameters used to
characterize $I$ because the latter are ill-defined for
non-positive-definite matrices such as ${\cal D}$,
for which the trace and the single eigenvalues may vanish or be negative.}
Certain correlations are detected for some of these ratios, meaning
that the degree of asymmetry between the principal axes of the tidal field
plays a role in determining the Lagrangian shape of a proto-halo.
For instance, it appears that the ratio $d_2/d_1$ is anti-correlated
with $i_3/i_2$.
This may be interpreted as another face of the $T$-$I$ correlation (\se{T-I}),
where the largest (smallest) compression flow in a proto-halo is
associated with its major (minor) axis of inertia.
For example, when the two largest compression flows are of comparable strength,
$d_1 \sim d_2 > d_3$, the proto-halo configuration is oblate,
$i_3 < i_2 \sim i_1$, namely a large $d_2/d_1$ and a small $i_3/i_2$.
On the other hand, when the compression is mainly along one
axis and $d_2 \simeq 0$ (i.e. small $d_2/d_1$), the configuration is prolate,
$i_3 \sim i_2 < i_1$ (i.e. large $i_3/i_2$).
Note that these correlations, much like the $T$-$I$ correlation,
derive from the constraint that
the Lagrangian region is destined to end up in a virialized halo.
When pancakes or filaments form instead, the expected correlations
may be
different, with the final minor axis of inertia correlated
with the direction of the largest flow of compression, etc.
As seen in \fig{hist}, the typical evolution from an initial proto-halo to a
final halo indeed passes through an intermediate phase of a pancake
(or a filament) lying perpendicular to the direction(s) of first collapse.

Back to proto-haloes, it is clear that binary correlations cannot tell
the full story; for a complete study one should consider the
joint distribution of the three eigenvalues.  Other quantities, such as
bulk flows, may also influence the shapes of proto-haloes.
Therefore, this issue deserves a more detailed study beyond the scope
of the current paper, which only provides first clues.

From the dynamical point of view,
proto-haloes can also be classified by the signs of the eigenvalues
of the deformation tensor at the centre of mass, indicating whether the initial
flows along the principal directions are of compression of dilation.
We find that a small minority of the proto-haloes, about 11 per cent,
are contracting
along all three principal axes.  The vast majority, about 86 per cent, are
initially collapsing along two directions and expanding along the third.
Only 2.7 per cent are collapsing along one direction and expanding along
two.\footnote{These fractions are
obtained with standard top-hat smoothing of the deformation tensor.
Similar numbers (15, 83, 1.5 per cent) are found when only the external
velocity field is taken into account
and minimal smoothing is applied ({\it Method 3} of Paper I).
However, when the deformation tensor is computed with minimal smoothing
and includes the contribution of fluctuations inside the proto-halo
({\it Method 2}), the frequencies become 98.3, 1.7, 0.0 per cent,
namely almost all the proto-haloes are collapsing along three spatial
dimensions. This is because the local gravitational attraction towards
the centre is dominant over the external shear such that it turns the
large-scale expansion associated with $d_3$ (and $i_3$) into a local
contraction.
This could be a feature that distinguishes proto-haloes from random
patches.
}
This clearly distinguishes the centres of mass of proto-haloes from
random points in the Gaussian field, where
the probability of the corresponding dynamical configurations
based on the deformation tensor would have been
8, 42, and 42 per cent respectively, with the rest 8 per cent expanding
along three directions (Doroshkevich 1970).
It also distinguishes the proto-haloes from peaks of the linear density
field. When smoothed with a top-hat window corresponding to 100 particles,
we find
in our simulation that the probabilities of the three dynamical
configurations are 45, 46 and 8.6 per cent, respectively.
These fractions
somewhat
depend on the smoothing length. Using a top-hat window
containing 1000 (10000)
particles, the corresponding probabilities become 50 (67), 44 (30) and 5.4
(3.4) per cent.
The peaks are characterized by compression along three or two directions,
while most of the proto-haloes are compressing along two directions.
This finding is consistent with the picture arising from \se{gamma}
where proto-haloes are typically embedded in elongated, filament-like
large-scale structures, surrounded by voids that induce compression
flows along the two principal directions orthogonal to the filament.

\begin{figure}
\epsfxsize= 8 cm \epsfbox{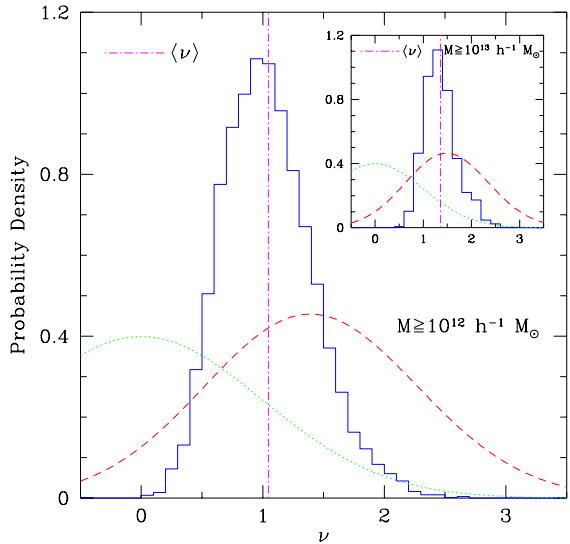}
\caption{
Proto-haloes versus peaks:
Distribution of density height $\nu =\delta /\sigma$ at proto-halo centres
(solid histogram)
versus random points (dotted line) and density maxima (dashed line).
The density is smoothed top-hat over the proto-halo scale.
The predicted probability distributions for density maxima and for
random points are for a corresponding linear Gaussian overdensity field,
smoothed on the scale of the smallest halo in the sample.
The main panel refers to all the haloes containing
more than 100 particles, while the inset refers to
haloes of more than 1000 particles.
The analytic estimates for peaks practically coincide with the actual
density distribution of peaks in our simulation.
The averages values, of $1.05$ and $1.35$, are marked by the dot-dashed lines.
The corresponding standard deviations are
0.37 and 0.33, respectively.
}
\label{fig:pkdist}
\end{figure}

\begin{figure}
\epsfxsize= 8 cm \epsfbox{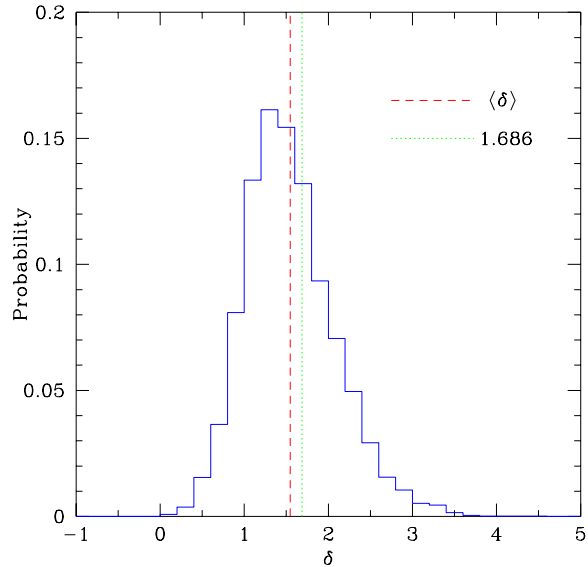}
\caption{
Comparison to the spherical collapse model:
Distribution of overdensity for proto-haloes, evaluated at the centre of
mass, smoothed top-hat on the halo scale, and linearly extrapolated
from the initial conditions to $z=0$.
The average value of the distribution, $\langle \delta \rangle=1.55$,
is marked by the dashed line.
The standard deviation of the distribution is 0.54.
The value predicted by the spherical collapse model, $\delta \simeq
1.686$, is marked by the dotted line.
For haloes with more than $1000$ particles,
the average is 1.47 and the standard
deviation is 0.36.
}
\label{fig:dedist}
\end{figure}

A certain fraction of
our proto-haloes are associated, to some degree,
with density peaks in the initial conditions.
In \fig{pkdist} we show the distribution of
$\nu=\delta/\sigma$
over all the haloes
containing more than 100 (main panel) or 1000 (inset) particles, 
where $\delta$ is
the density contrast at the centre of mass of the proto-halo and $\sigma$
is the rms density contrast over all space.
This is compared with the expected distribution of $\nu$,
for random points and for density peaks,
in a Gaussian random field with the $\tau$CDM power spectrum, smoothed on the
scale corresponding to 100 (main panel) and 1000 (inset) particles
of our simulation.
Note that our proto-haloes typically correspond to one-$\sigma$ fluctuations
with a relatively small
dispersion.
This distribution can be clearly
distinguished from the density distribution in randomly selected density
maxima, which average at about $1.4\sigma$, and have extended tails --
especially towards high values.
Some of the high peaks are embedded in more extended perturbations
of $\nu \sim 1$.
They give rise to clumps that eventually
merge into the larger haloes that we identify today, and they are
therefore not included in our sample of proto-haloes.
Also, recall that the impression from \fig{ex1} and \fig{ex2} was that
the correlation between the halo centres of mass and the positions of
density peaks is quite limited,
and therefore the central proto-halo densities tend
to be lower than the heights of the associated peaks.
It is also possible that
some very high and very low peaks may have deformation configurations that
would end up as very aspherical or fragmented structures rather than
coherent virialized haloes
(e.g. Katz, Quinn \& Gelb 1993; van de Weygaert \& Babul 1994).

One often uses the terminology of the
spherical collapse model to characterize the evolution of proto-haloes.
For instance, one refers to `turnaround' and `collapse' time, even though,
strictly speaking, these
quantities are not well-defined for non-spherical objects
(see, however, the discussion in Sugerman \etal 2000).
One way to test our definition of just-collapsed haloes at $z=0$,
and to address the sphericity of proto-haloes, is by evaluating
the accuracy with which the spherical collapse model describes
their evolution.
\fig{dedist} shows the distribution of the linearly extrapolated density
contrast at the proto-halo centre of mass, smoothed on the halo size.
The average of the distribution, $\delta=1.55$,
is within $\sim$10 per cent of the standard prediction
of the spherical collapse model at collapse time, $\delta\simeq
1.686$.
This indicates that our halo finding method and the spherical model are
consistent on average.
However, the scatter is large, indicating strong deviations from sphericity
for many individual proto-haloes, which implies a big uncertainty in the
turnaround or collapse times.

What are the proto-haloes that populate the tails of the linear overdensity
distribution? In particular, how do proto-haloes of $\delta \ll 1$ manage
to make haloes by today?
It is expected that the presence of shear may speed-up the collapse
of a density perturbation with respect to the spherical case
(Hoffman 1986b; Zaroubi \& Hoffman 1993; Bertschinger \& Jain 1994).
In fact, our haloes with $\delta \ll 1$ are indeed characterized by a strong
shear. They typically have $d_1\simeq -d_3$ and $|d_2|\ll |d_3|$,
i.e. they are characterized by compression and dilation factors
of similar amplitudes on orthogonal axes.
In this case, shear terms in the Raychaudhuri equation
(e.g. Bertschinger \& Jain 1994) account for significant corrections
to the spherical terms, already at $z=50$.
On the other hand, there are high density proto-haloes, $\delta \sim 3$, which
collapse only today, which is late compared to the predictions of the
spherical model.  These are generally characterized by strong compression flows
along 2 directions and by mild expansion along the third axis.
The reason for their late collapse is not obvious
and deserves further investigation which is beyond our scope here.

\section{Alignment of Spin and Shear}
\label{sec:T-L}

\begin{figure}
\epsfxsize= 8 cm \epsfbox{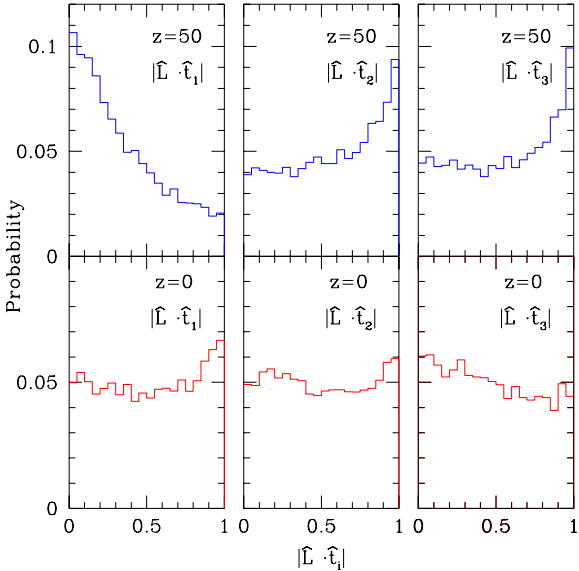}
\caption{
Correlation between halo spin and the linear shear tensor.
Shown is the distribution of the cosine of
the angle between the
halo spin (at $z=50$ or at $z=0$) and each of the three eigenvectors of
the linear shear tensor.
The averages and standard deviations at $z=50$, for axes 1,2,3 respectively,
are:
$0.35\pm0.27$, $0.56\pm0.30$, $0.55\pm0.31$.
The corresponding values at $z=0$ are:
$0.52\pm0.30$, $0.50\pm0.30$, $0.47\pm0.29$.
}
\label{fig:L_vs_T}
\end{figure}

Had the inertia and shear tensors been {\it uncorrelated}, one could have
argued that the direction of the spin in the linear regime should
tend to be aligned with the direction of the middle eigenvector of $T$ (and of
$I$).
The Cartesian components of $L$ in the frame where $T$ is diagonal are
\begin{equation}
L_i\propto (t_j-t_k)I_{jk} \,,
\end{equation}
where $i,j,k$ are cyclic permutations of $1,2,3$.
We average over all the possible orthogonal matrices $R$ that may relate
the uncorrelated principal frames of $T$ and $I$, while keeping the
eigenvalues of the two tensors fixed, and obtain
\begin{equation}
\langle |L_i| \rangle \propto |t_j-t_k| \,
\langle\, |I_{jk}|\ |i_1,i_2,i_3 \rangle
= |t_j-t_k| \, f(i_1,i_2,i_3)\, .
\end{equation}
The assumed independence of $T$ and $I$ would have guaranteed that in the
principal frame of $T$ the conditional average of $I_{jk}$
(given the eigenvalues of $I$) would have been the same for all $j \neq k$,
namely a certain function $f(i_1,i_2,i_3)$.
On average, the largest component $|L_i|$ would have been the one for which
$|t_j-t_k|$ is the largest, which is necessarily $L_2\prop|t_3-t_1|$,
because by definition $t_1\leq t_2 \leq t_3$.  Thus, the angular momentum
would have tended to be aligned with ${\bf t}_2$.\footnote{Note
that a similar argument can be used to show that $L$ would have also been
preferentially aligned with the middle eigenvector of $I$.
In this case, however, since all the eigenvalues $i_j$ are positive, one
expects a weaker correlation.
}
To demonstrate this argument and evaluate the strength of the effect,
we use the randomized
$T$-$I$ pairs obtained as described at the end of \se{T-I}. In this case,
we find
$\la \hat{\am}\cdot \hat{\bf t}_i \ra = 0.40, 0.66, 0.38$ (for $i=1,2,3$),
with a scatter of $\sim 0.3$ about these mean values.
We see that the tendency for alignment with the middle eigenvector is
quite significant.

Given that the inertia and shear tensors are in fact strongly {\it correlated}
(\se{T-I}), we now investigate, using TTT, what kind of alignment, if any,
one should expect between $\am$ and $T$ at the initial conditions.
Due to the $T-I$ correlation, the distribution of the orthogonal linear
transformations $R$ should not be uniform but rather favour rotations with
small Euler angles.  Moreover, since we saw in \se{T-I} that different pairs
of eigenvectors have different degrees of alignment, the average off-diagonal
terms of $I$ in the principal frame of $T$ would not be all equal anymore.
This could be sufficient for drastically changing the preferential
alignment of the angular momentum with ${\bf t}_2$ seen in the uncorrelated
case.
Using TTT based on the inertia and tidal tensors for proto-haloes as extracted
from the simulation, we actually obtain
$\la \hat{\am}\cdot \hat{\bf t}_i \ra = 0.34, 0.59, 0.52$,
with a scatter of $\sim 0.3$ in every case.
This shows that the $T$-$I$ correlation indeed modifies
the $\am$-$T$ alignment predicted in the uncorrelated case.
The spin tendency for alignment with ${\bf t}_2$ is still the strongest,
but it is weaker than before. The alignment with ${\bf t}_3$ becomes
stronger and almost comparable to the alignment with ${\bf t}_2$,
while the alignment with ${\bf t}_1$ becomes even weaker than it was,
namely the spin prefers to be perpendicular to ${\bf t}_1$.
We see that, generally speaking, the correlation between $I$ and $T$
tends to weaken the correlation between $\am$ and $T$.

Next, we measure the actual spins of the proto-haloes in the initial conditions
of the simulation, and correlate it with the directions of the eigenvectors of
$T$.
The distributions of angles between these vectors are shown in \fig{L_vs_T}
(top panels). We see that the spin has a weak but significant tendency
to be perpendicular to the major axis of $T$, and an even weaker tendency
to be parallel either to the middle or to the minor axis of $T$.
In this case,
$\la \hat{\am}\cdot \hat{\bf t}_i \ra = 0.35, 0.56, 0.55$,
in good agreement with the TTT approximation.

Thus, contrary to LP00, we find no evidence for a strong
correlation with the direction of the middle axis of $T$.
LP00
have proposed to parametrize $\am$-$T$ correlations through the expression
\begin{equation}
\label{eq:LP}
\langle \hat{L}_i \hat{L}_j | \hat{T}_{ij} \rangle=
\f{1+a}{3} \delta_{ij}-a \hat{T}_{ik}\hat{T}_{kj}\;,
\end{equation}
where $\hat{T}_{ij}=T_{ij}/[T_{lk}T_{kl}]^{1/2}$ is the trace-free unit
shear tensor, and $a$ is the correlation parameter.
\footnote{We
notice that the claim of LP00, that \equ{LP} represents the most
general quadratic relation between a unit vector and a unit tensor,
is not formally true, because, for example, \equ{LP} cannot describe
the case where $\hat{\am}$ systematically lies along one of
the principal axes of $T$.
However, since this configuration cannot be realized in TTT, we
adopt \equ{LP} as a plausible parametrization of the correlation.}
This is particularly convenient for analytical studies of spin statistics
(e.g. Crittenden \etal 2001), since it bypasses the ambiguous issue
of determining the inertia tensors of proto-haloes and their statistics.
The correlation parameter $a$ vanishes if $\am$ is randomly distributed
with respect to $T$, and according to LP00 it should take the value
$\sim 3/5$ if TTT holds and $T$ is independent of $I$
(empirically, we actually get $a=0.68\pm 0.01$ from the randomized
sample in our simulation).
It is especially interesting to provide an independent measure of
$a$ from our simulation because \equ{LP} has been used in modeling spin-spin
correlations (Crittenden \etal 2001) and only one determination of
$a$ is available (LP00).

The correlation parameter $a$ can be evaluated in the principal frame
of the tidal tensor.  Denoting the eigenvalues and eigenvectors of $\hat{T}$
by $\tilde {t}_i$ and ${\hat{\tilde{\bf t}}} _i$,
one obtains
$a=2
-6\langle\sum_i(\hat{\am}\cdot \hat{\tilde{{\bf t}}}_i)^2 \,
\tilde {t}_i^2\rangle$,
where $\sum \tilde{t}_i^4=1/2$ has been used (LP00).
Using this expression for $a$, we obtain by averaging over
all the proto-haloes in our initial conditions $a=0.28\pm 0.01$,
and when we restrict ourselves to proto-haloes containing 1000
particles or more we get $a=0.33\pm 0.04$.
It is worth stressing, however, that the dispersion among the proto-haloes
of the quantity that is averaged in the determination of $a$
is very large, with an rms value of about $0.9$,
which is not a good property of this statistic.

Next, we wish to find out how much of the (weak) ${\bfL}$-$T$ correlation
detected at the initial
conditions actually survives the non-linear evolution to the present epoch.
For this we plot in the bottom panels of \fig{L_vs_T} the distributions
of angles between the {\it final\,} spin and the initial shear tensor.
We find that the correlation is 
strongly weakened.
This is hardly surprising based on the significant non-linear
evolution in spin direction found in Paper I.
We obtain $a=0.07\pm 0.01$, at variance with
the result obtained by LP00, $a=0.24\pm 0.02$
at $z=0$,
using a low-resolution particle-mesh simulation of an $\Omega=1$ CDM model.
Our result persists when considering only the haloes containing
1000 particles or more,  $a=0.07\pm 0.04$.
The discrepancy with LP00 is probably an effect of the higher accuracy with
which our high-resolution, adaptive P$^3$M code accounts 
for non-linear effects.
The smoothing length used to define $\hat T_{ij}$ in the simulation
could also affect the results
(for instance, smoothing on too small a scale might give unphysically
small $L-T$ correlations).
In order to test 
the dependence of our findings on the adopted filtering radius,
we re-compute $a$ using a top-hat window function which contains 
8 times the halo mass. In this case, for haloes with 100 particles or more,
we find $a=0.24\pm 0.01$ at z=50, and $a=0.08\pm 0.01$ at $z=0$.
This shows that smoothing is not a major issue here. 

Our result above seems to suggest that observed spin directions,
as deduced from disc orientations,
cannot be used to deduce useful information about the initial shear tensor.
How relevant are our results to the orientation of disc galaxies?
On one hand, the haloes in our current analysis are all characterized
by a present-day overdensity of $\sim 180$, while the galactic discs
are associated with much higher density contrasts, so a certain
unknown extrapolation is required.
Moreover, it is plausible that the plane of today's disc galaxy
is determined by the halo spin at an early time, before the gas fell into
the disc.  As expected, we see in the simulation that the non-linear effects
wipe out the spin-shear correlation in a gradual fashion, e.g.,
$a=0.26, 0.20, 0.11$ ($\pm 0.01$) at $z=1.2, 0.6, 0.2$ respectively.
Therefore, the disc orientation may after all preserve some of the
weak correlation with the initial shear field, and this should be true
especially for galaxies that formed at high redshift.
On the other hand, we know that the spin and shear are correlated better
at high redshift only for the haloes that we selected at $z=0$, and we
do not know whether this is true for the subhaloes which host
disc formation at higher redshift.
It is also possible that the statistical orthogonality between
the spin and the first principal shear axis is preserved
better in specific regions were cosmic flows are particularly cold.
The cold-flow neighborhood of the Local Group might provide
a suitable field for studying such a correlation.
Still, we expect only a weak correlation.

\section{Conclusion}
\label{sec:conc}

In order to deepen our understanding of how tidal torques actually work,
we investigated the cross-talk between the main components of the theory
using $\sim 7300$ well-resolved haloes extracted from a cosmological
$N$-body simulation.
We studied the correlation between the proto-halo inertia tensor
and the external shear tensor, and between them and the spin direction,
and tried to characterize the proto-halo regions in several different ways.
We defined haloes in today's density field using the standard
friends-of-friends algorithm,
but it would be useful for future work to check
robustness to alternative halo finders as well as to different
cosmological scenarios.

We found to our surprise that the $T$ and $I$ tensors are strongly correlated,
in the sense that
their minor, major and middle principal axes tend to be aligned, in this order.
This means that the angular momentum, which plays such a crucial role
in the formation of disc galaxies, is only a residual which arises
from the little, $\sim$10 per cent
deviations from perfect alignment of $T$ and $I$.

The $T-I$ correlation induces a weak tendency of the proto-halo spin
to be perpendicular to the major axis of $T$ (and $I$). It also slightly
weakens the spin tendency to be aligned with the middle axis
(a tendency that would have dominated had $T$ and $I$ been uncorrelated)
and slightly strengthen its alignment with the minor axis.
However, these correlations, even at the initial conditions, are weak.
Furthermore, non-linear changes in spin direction at late times
practically erase the memory of the initial shear tensor, and therefore
observed spin directions cannot serve as very useful indicators
for the initial shear tensor (again, in variance with LP00).
The only partial caveat is that today's discs may reflect the halo
spin directions at some high redshift, which may still preserve
some weak correlation with the linear shear field.
A study of this effect
would require a high-resolution simulation in which a detailed galaxy
formation scheme is incorporated.

On the other hand, the strong $T-I$ correlation provides a promising hint
for how to make progress in a long-standing open question in galaxy
formation theory. That is, how to characterize proto-haloes and their
boundaries.  Speculations based on iso-density or iso-potential contours
about high-density peaks have failed the tests of simulations.
We first realize that the centres of proto-haloes tend to lie in
$\sim 1\sigma$ over-density regions, but their association with the
linear density maxima smoothed on galactic scales is quite limited.
We find that the proto-haloes tend to be elongated along the direction
of maximum compression of the smoothed velocity field.
A typical configuration is of an elongated proto-halo lying
perpendicular to an elongated background density ridge, with neighboring
voids inducing the compression along the major and intermediate inertia axes.
The $T-I$ correlation detected here should thus enable the
construction of a detailed algorithm to identify 
proto-haloes and their boundaries 
in cosmological initial conditions
(Porciani, Dekel \& Hoffman, in preparation).

The collapse proceeds in two succesive stages, corresponding to the
quasi-linear and the non-linear regimes.  The first stage is
characterized by a two dimensional collapse in the plane defined by
the major and intermediate inertia axes.  This leads
to the formation of a transient clumpy elongated structure aligned
with the large scale filament within which the halo resides.  In the
fully non-linear regime the proto-halo experiences a one-dimensional
collapse and a series of merger of the sub-clumps along the filament,
leading to a quasi-spherical object in virial equilibrium.

The proto-haloes can be characterized by a variety of other statistics,
which may distinguish them from over-density patches about random density
peaks.
In terms of shape, most proto-haloes have significant ellipticities and
most of them are prolate,
but this
does not distinguish them very clearly from over-density patches
about peaks.
In contrast with the strong alignment of the inertia and deformation tensors,
there are
only
weak correlations between the triaxiality of the inertia tensor
and that of the deformation tensor,
as expressed by ratios of eigenvalues.

The vast majority of proto-haloes are initially collapsing along two
principal directions
(except when the self gravity is also considered, inducing a collapse
in all directions).
This is in clear contrast with the behaviour of density peaks, which
have similar probabilities to collapse along three or two axes.
Our finding is consistent with proto-haloes being
embedded in large-scale filaments surrounded by voids.
The smoothed central densities in proto-haloes
typically represent $1\sigma$ positive perturbations, but with
smaller dispersion than for general peaks. This is partly because
the proto-halo centers are only weakly correlated with the density peaks,
and also because some of
the high peaks lead to clumps that later merge into the big haloes we identify
today and include in our sample.
Despite the significant asphericity,
the proto-halo densities agree on average with the predictions of the
spherical collapse model, but the scatter is large.
For example, there are proto-haloes with
$\delta \ll 1$ (with $\delta$ the density contrast smoothed on the
halo scale and linearly estrapolated to today), whose collapse
is boosted by strong shear of large compression and dilation factors.

In Paper I, we evaluated the performance of linear TTT in predicting
the spin of galactic haloes.
We found that, for a given proto-halo at the initial conditions,
TTT provides a successful order-of-magnitude estimate of the final halo
spin amplitude. The TTT prediction matches on average the spin amplitude
of today's virialized
haloes if linear TTT growth is assumed until about $t_0/3$.
This makes TTT useful for studying certain aspects of galaxy formation,
such as the origin of a universal spin profile in haloes (Bullock \etal 2001b;
Dekel \etal 2001), but only at the level of average properties, because
the random error is comparable with the signal itself.
We also found in Paper I that
non-linear evolution causes significant variations in spin direction,
which limit the accuracy of the TTT predictions to a mean error of
$\sim 50^\circ$. Furthermore, spatial spin-spin correlations
on scales $\geq 1\hmpc$ are 
strongly weakened by non-linear effects.
This limits the usefulness of TTT in predicting intrinsic galaxy alignments
in the context of weak gravitational lensing (Catelan \etal 2001; Crittenden
\etal 2001).
This situation may improve (as in the case of the spin-shear correlation
addressed in the current paper) if
the orientations of today's discs were determined by the halo spins at
a very high redshift, which are better modelled by TTT.
On the other hand, we know this only for the haloes that we selected at
$z=0$, and we don't know whether this is true for the subhaloes which host
disc formation at higher redshift.

In our studies of the cross-talk between the different components of TTT,
we have also
realized
another surprise that leads to a revision in
the standard scaling relation of TTT (White 1984).
It is demonstrated in Paper III that the off-diagonal, tidal terms
of the deformation tensor, which drive the torque, are 
only weakly 
correlated with the diagonal terms, which determine the overdensity at the
proto-halo centre.  The latter enters the TTT scaling relation via the expected
collapse time of the proto-halo, and the lack of correlation with the torque
leads to a modification in the scaling relation.
The revised scaling relation can be applied shell by shell
in order to explain the origin of the universal angular-momentum profile
of haloes (Bullock \etal 2001b; Dekel \etal 2001). It can also be very
useful in incorporating spin in semi-analytic models of galaxy formation
(Maller, Dekel \& Somerville 2002).

\section*{Acknowledgments}

This research has been partly supported by the Israel Science Foundation
grants 546/98 and 103/98,
and by the US-Israel Binational Science Foundation grant
98-00217.  CP acknowledges the support of a Golda Meir fellowship at HU
and of the EC RTN network `The Physics of the Intergalactic Medium' at the
IoA.  We thank our GIF collaborators, especially
H.  Mathis, A. Jenkins,
and S.D.M. White, for help with the GIF simulations.


\bsp
\end{document}